\pdfoutput=1
\documentclass{aastex701}

\usepackage{amsmath}

\begin{document}

\title{Disentangling the Morphology of Palomar~5: Effects of the Bar, Spiral Arms, LMC, and Halo Flattening}


\correspondingauthor{Long Wang}
\email{wanglong8@sysu.edu.cn}

\author[orcid=0009-0005-2126-3824]{Zhenghao He}
\affiliation{School of Physics and Astronomy, Sun Yat-sen University, Daxue Road, 519082 Zhuhai, China}
\email{hezhh59@mail2.sysu.edu.cn}

\author[orcid=0000-0001-8713-0366]{Long Wang}
\affiliation{School of Physics and Astronomy, Sun Yat-sen University, Daxue Road, 519082 Zhuhai, China}
\affiliation{CSST Science Center for the Guangdong-Hong Kong-Macau Greater Bay Area, 519082 Zhuhai, China}
\email[show]{wanglong8@sysu.edu.cn}

\author[orcid=0009-0009-7509-3794]{Zi-yi Zhou}
\affiliation{School of Physics and Astronomy, Sun Yat-sen University, Daxue Road, 519082 Zhuhai, China}
\email{zhouzy83@mail2.sysu.edu.cn}

\author[orcid=0000-0003-3250-2876]{Yang Huang}
\affiliation{School of Astronomy and Space Science, University of Chinese Academy of Sciences, Beĳing 100049, China}
\affiliation{National Astronomical Observatories, Chinese Academy of Sciences, Beijing 100101, China}
\email{huangyang@ucas.ac.cn}

\author[orcid=0000-0002-5038-9267]{Eugene Vasiliev}
\affiliation{University of Surrey, Guildford, GU27XH, United Kingdom}
\email{eugvas@protonmail.com}


\begin{abstract}
The Palomar~5 (Pal~5) globular cluster and its tidal tails provide a sensitive probe of globular-cluster evolution in the time-dependent Milky Way potential. We study the past 3~Gyr evolution of Pal~5 using collisional direct \(N\)-body simulations with \texttt{PeTar}, adopting Galactic potential models that include spiral arms, the Galactic bar, halo flattening, the Large Magellanic Cloud (LMC), and bar deceleration. We find that halo shape strongly influences the projected stream track, reflecting Pal~5's sensitivity to Galactic force-field flattening. The LMC causes only modest direct changes to the present-day projected stream morphology, but can alter the pericentric distance and hence the progenitor's mass evolution. The Galactic bar strongly affects stream length and debris redistribution along the tails, producing model-dependent density structures and leading--trailing asymmetries. Comparison with observations from \citet{Erkal2017} and \citet{Xiao2025} shows that no single model simultaneously reproduces all observed properties of Pal~5, including cluster evolution, stream length, track, width, and line-density profile. Although our simulations capture several global properties, the remaining discrepancies indicate that a more precise match likely requires better constraints on the initial properties of the Pal~5 progenitor and a more complex Galactic potential, including perturbations from small-scale perturbers such as dark matter subhalos and giant molecular clouds. Future work may combine self-consistent direct \(N\)-body simulations with particle-spray methods to investigate these discrepancies more efficiently.
\end{abstract}



\section{Introduction} \label{sec:intro}

The shape and time dependence of the Milky Way (MW) gravitational potential are key
uncertainties in Galactic dynamics. Measurements of the Galactic force field provide
constraints on the distribution of baryons and dark matter, the response of the halo
to satellite accretion, and the dynamical influence of non-axisymmetric structures
such as the Galactic bar and spiral arms \citep[e.g.,][]{VeraCiro2013, Bovy2016,
Han2023, Nibauer2025}. Stellar streams are particularly useful probes of this
potential because their orbits, widths, and density structures preserve information
about both the global force field and local gravitational perturbations
\citep{Johnston1999, Helmi2008, Ibata2001, Koposov2010}. Streams produced by
disrupting dwarf galaxies trace the large-scale halo potential, while the thin and
dynamically cold streams formed from globular clusters are especially sensitive to
small-scale changes in the Galactic acceleration field and to perturbations from compact
or time-dependent structures \citep{Baumgardt2003, Yoon2011, Carlberg2012,
Erkal2015, Erkal2016, Bonaca2019}.

Among the known globular-cluster streams, the Palomar~5 (Pal~5) stream is a
particularly important case. Pal~5 has a surviving, low-mass progenitor near the
apocentre of its orbit, and its prominent S-shaped tidal tails extend for more than
\(20^\circ\) across the sky \citep{Odenkirchen2001, Rockosi2002, Grillmair2006}.
Because the stream is thin and dynamically cold, it has long been used as a precise
probe of the Galactic acceleration field in the inner halo
\citep{Kupper2015, Ibata2016}. 
The observed leading--trailing asymmetry, density variations, and
possible gaps provide additional constraints on both the Galactic potential and the
dynamical history of the progenitor. Some density structures may arise naturally
from internal stream dynamics, such as epicyclic overdensities
\citep{Kupper2008, Kupper2010, Kupper2012}, while others may be produced by external
perturbations, including dark matter subhalos \citep{Yoon2011, Carlberg2012,
Erkal2015, Erkal2016}, giant molecular clouds \citep{Amorisco2016}, or flybys of
other globular clusters \citep{Ferrone2025}. For Pal~5, \citet{Erkal2017} and
\citet{Bonaca2020} showed that the stream contains significant density variations
whose origin is not unique. More generally, non-axisymmetric structures such as the
Galactic bar and spiral arms can reshape cold streams and generate density features
that may resemble the signatures of other perturbers
\citep{Hattori2016, Pearson2017, Starkman2020}.

Accurately identifying the dominant mechanisms behind the Pal~5 morphology requires
models that combine a realistic treatment of the disrupting cluster with a
time-dependent Galactic potential. Many previous stream studies have relied on
static potentials or approximate collisionless techniques, such as particle-spray
or streakline methods \citep{Kupper2012, Fardal2015, Chen2025}. These approaches are
computationally efficient, but they generally approximate the stripping process and
do not self-consistently follow the collisional internal evolution of the progenitor.
This is important for Pal~5, whose mass loss and stream formation are directly tied
to the internal evolution of the cluster \citep{Gieles2021, Wang2024, Roberts2025}.
At the same time, the MW potential itself is not static. The Galactic bar and spiral
arms can drive non-axisymmetric resonances \citep{Hattori2016, PriceWhelan2016}, the
bar may decelerate due to dynamical friction with the dark matter halo
\citep{Chiba2021_Tree, Dillamore2024, Zhang2025}, and the infall of the Large
Magellanic Cloud can induce reflex motion and distort the MW halo
\citep{Erkal2019, Erkal2021, Petersen2021, GaravitoCamargo2021, Vasiliev2021,
Shipp2021, Lilleengen2023, Koposov2023, Brooks2025}. Recent cosmological stream
frameworks further show that time-evolving Galactic potentials can naturally produce
stream clumps, orbital-plane evolution, and phase-dependent track misalignments
\citep{Panithanpaisal2026}. However, coupling such fully evolving environments to
self-gravitating, collisional cluster simulations remains computationally expensive.
Direct collisional \(N\)-body simulations embedded in tailored, time-dependent
analytic potentials therefore provide a complementary approach: they do not capture
the full cosmological evolution of the MW, but they allow us to isolate how selected
components, such as the bar, spiral structure, halo shape, the LMC, and bar
slowdown, affect both the bound cluster and its tidal stream.

In this work, we perform collisional direct \(N\)-body simulations of Pal~5 using
the \texttt{PeTar} code \citep{Wang2020_PeTar}. Using the $N$-body model of \cite{Wang2024} as a baseline, we introduce a time-dependent MW potential with variations in the rotating bar, spiral arms, halo flattening, and the presence of the LMC 
to examine how different Galactic components affect the evolution
of the bound cluster and the morphology of its tidal stream. This paper is organized
as follows. Section~\ref{sec:methods} describes the numerical method, including the
\texttt{PeTar} code, the Galactic potential models, and the initial conditions of
the Pal~5 progenitor. Section~\ref{sec:results} presents the simulation results,
including the cluster mass evolution, structural evolution, orbital properties, and
the morphology of the tidal stream. We then compare the simulated streams with
observational measurements from \citet{Erkal2017} and \citet{Xiao2025}, focusing on
the stream length, projected track, transverse width, and line-density profile.
Finally, Section~\ref{sec:discussion_conclusion} summarizes the main results and
discusses the limitations of the present models and possible directions for future
work.


\section{Methods} \label{sec:methods}


\subsection{N-body code} \label{sec:nbody}

To simulate the dynamical evolution of Pal~5, we use \texttt{PeTar} \citep{Wang2020_PeTar}, a highly optimized direct $N$-body code designed for dense stellar systems. \texttt{PeTar} is built upon the Framework for Developing Parallel Particle Simulation Codes (FDPS; \citealt{Iwasawa2016, Iwasawa2020}), which allows for efficient parallelization across large computing clusters.

The core algorithm of \texttt{PeTar} is based on the Particle-Particle-Particle-Tree (P$^3$T) scheme \citep{Oshino2011}, which divides gravitational interactions into long-range (``soft'') and short-range (``hard'') components. The integration cycle follows a Hamiltonian splitting approach:
\begin{itemize}
    \item \textbf{Soft Step (Long-range):} Global gravitational forces are computed using the \citet{Barnes1986} tree method combined with a symplectic second-order Leap-frog integrator. To maximize performance, the soft force kernels are accelerated using SIMD instructions (e.g., AVX512) or GPU (CUDA) technologies, achieving a computational complexity of $\mathcal{O}(N \log N)$.
    \item \textbf{Hard Step (Short-range):} For particles in high-density regions, binaries, or few-body subsystems, \texttt{PeTar} identifies neighbor lists and constructs local binary trees. These interactions are integrated using a fourth-order Hermite predictor-corrector scheme. Crucially, high accuracy for close encounters and binary orbits is maintained via the Slow-Down Algorithmic Regularization (SDAR) method \citep{Wang2020_SDAR}, which prevents the time-steps from becoming too small during hard binary evolution.
\end{itemize}

For this study, a key feature of \texttt{PeTar} is its ability to embed the $N$-body system within an external, time-dependent tidal field. We couple the cluster's internal dynamics with Galactic potential models, allowing us to account for both internal self-gravity and external tidal forces under realistic Galactic conditions.
Although some $N$-body codes have an option to provide an external tidal field, it is either simplified (e.g. a constant tidal field corresponding to a cluster on a circular orbit), or requires the user to write their own potential function \citep{Renaud2015}. In contrast, \texttt{PeTar} interfaces with two general-purpose galactic dynamics libraries, \texttt{Galpy} \citep{2015ApJS..216....9B} and \texttt{AGAMA} \citep{Vasiliev2019}, which provide flexible potentials with adjustable parameters. \texttt{AGAMA}, recently implemented in \texttt{PeTar} and widely used in other codes including \texttt{Arepo} \citep{Springel2010}, \texttt{Gadget-4} \citep{Springel2021}, \texttt{NEMO} \citep{Teuben1995} and \texttt{AMUSE} \citep{PortegiesZwart2013}, supports arbitrary time-dependent potentials, such as a rotating bar with evolving pattern speed,  amplitude, and size. The time-dependent potential used in this work is implemented with \texttt{AGAMA}.


\subsection{Galactic potentials} \label{sec:potential}


\begin{deluxetable}{lcccc}
\tablecaption{Summary of the Galactic potential models and their specific configurations used in this work.\label{tab:potential_models}}
\tablehead{
\colhead{Model Name} & \colhead{DM Halo Shape} & \colhead{Galactic Bar} & \colhead{Spiral Arms} & \colhead{LMC}
}
\startdata
MW2014   & Spherical & None           & No  & No  \\
SHT2024  & Spherical & Constant Speed & Yes & No  \\
FHT2024  & Flattened & Constant Speed & Yes & No  \\
FNHT2024 & Flattened & None           & Yes & No  \\
FLHT2024 & Flattened & Constant Speed & Yes & Yes \\
FDHT2024 & Flattened & Decelerating   & Yes & No  \\
\enddata
\tablecomments{Model naming convention: 'S' and 'F' denote Spherical and Flattened dark matter halos, respectively. 'N' indicates the absence of a Galactic bar, 'L' indicates the inclusion of the LMC, and 'D' indicates a Decelerating bar. The suffix 'HT2024' refers to the basis model from \citet{hunterTestingKinematicDistances2024}.}
\end{deluxetable}


To study the dynamical effects of non-axisymmetric structures (spiral arms, bar), halo geometry, LMC, and bar deceleration on Pal~5, we use six different Galactic potential models.
We adopt the MW Potential used for the Pal~5 simulations in \cite{Wang2024} as our baseline model (MW2014). This model uses the time-independent, axisymmetric
MW potential, \texttt{MWPotential2014}, from \citet{2015ApJS..216....9B}, implemented
through the \texttt{GALPY} interface in \texttt{PeTar}. it consists of a
spherical NFW dark matter halo, a Miyamoto-Nagai disk, and a power-law bulge with
an exponential cutoff. We use it as a simple axisymmetric reference to
assess how additional Galactic components affect the evolution of Pal~5 and its
tidal stream.

Other five models adopts modified versions of the time-dependent MW model of
\citet{hunterTestingKinematicDistances2024} implemented in \texttt{PeTar} through the \texttt{AGAMA} interface. It includes a
multi-component central region, stellar and gas disks, a rotating Galactic bar  and spiral-arms, and a dark matter halo. The relevant component parameters are summarized in Appendix~\ref{app:potential_parameters}. Below, we describe the component modifications applied in our simulations.

Previous studies indicate that the MW halo may be flattened, motivating the inclusion of a flattened halo potential in our simulations. For a direct comparison between the sphercial and flattened halos, we replace the spherical \citet{1969AN....291...97E} profile used in  \citet{hunterTestingKinematicDistances2024} with a modified Navarro–Frenk–White (NFW) profile.

\begin{figure}[htbp]
    \centering
    \includegraphics[width=0.8\textwidth]{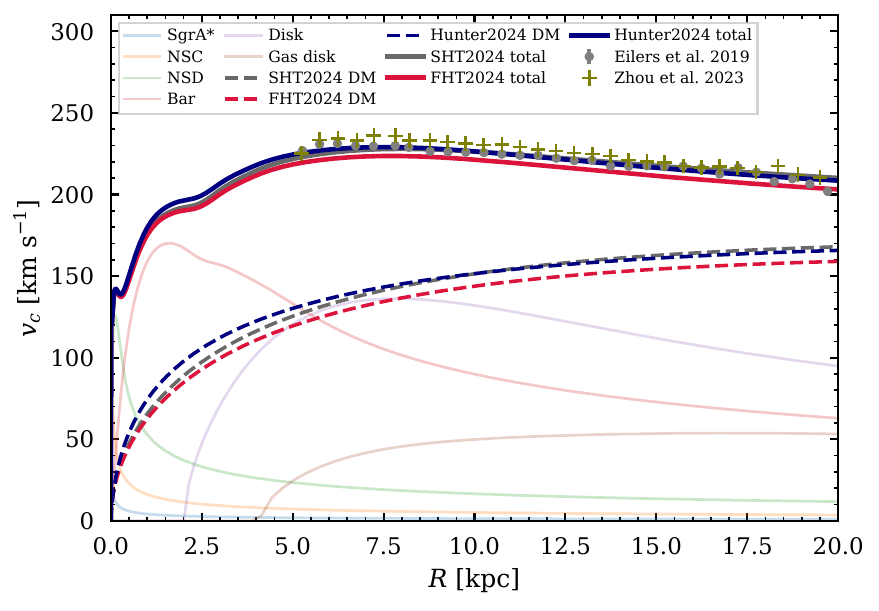}
    \caption{
         Circular-velocity curves for the SHT2024, FHT2024, and original Hunter2024 models. For each model, the total circular velocity is shown with a solid line, while the corresponding dark-matter halo contribution is shown with a dashed line of the same colour. The remaining potential components are plotted with transparent coloured lines. Observational measurements from \citet{Eilers2019} and \citet{Zhou2023} are included for reference. 
    }
    \label{fig:rotation_curve}
\end{figure}

The second model, SHT2024, adopts the standard NFW density profile \citep{1996ApJ...462..563N} is
\begin{equation}
    \rho_{\text{halo}}(r) = \frac{\rho_0}{(r/r_{\mathrm{s}})\,(1 + r/r_{\mathrm{s}})^2}, 
    \label{eq:NFW_rho}
\end{equation}
where $r = \sqrt{x^2 + y^2 + z^2}$ is the Galactocentric radius, $\rho_0$ is the characteristic density, and $r_{\mathrm{s}}$ is the scale radius. We adopt $\rho_0 = 1.21 \times 10^7~M_\odot~\text{kpc}^{-3}$ and $r_{\mathrm{s}} = 14.39~\text{kpc}$, consistent with the results of \citet{2016MNRAS.463.2623H}. This model gives a circular-velocity curve slightly below that of \citet{hunterTestingKinematicDistances2024} (Figure~\ref{fig:rotation_curve}), with a local circular velocity of $v_c = 228~\text{km\,s}^{-1}$ at the solar radius ($R_0 = 8.179~\text{kpc}$; \citealt{Gravity2019}), compared with the reference value of $229~\text{km\,s}^{-1}$ in \citet{hunterTestingKinematicDistances2024}.


The third model, FHT2024, introduces axisymmetric flattening into the modified NFW halo
by replacing the spherical radius \(r\) in Equation~\ref{eq:NFW_rho} with the
ellipsoidal radius
\begin{equation}
    m = \sqrt{x^2 + y^2 + \left(\frac{z}{q_{\rho}}\right)^2},
\end{equation}
here \(q_{\rho}\) is the minor-to-major axis ratio of the halo density
distribution. We adopt an oblate density halo with \(q_{\rho}=0.84\) as a
controlled flattened-halo test, designed to examine how halo non-sphericity affects the Pal~5 orbit and stream morphology. This density flattening also modifies the midplane radial force, making the FHT2024 circular-velocity curve slightly lower than the spherical SHT2024 curve (Figure~\ref{fig:rotation_curve}).

To assess the kinematic influence of the Galactic bar, we construct a fourth, control model, FNHT2024, by removing the bar potential from FHT2024. The triaxial rotating bar is replaced with its fully axisymmetric equivalent by truncating the azimuthal multipole expansion of the central bar-disk component at $m=0$. This preserves the overall radial mass distribution and the circular-velocity curve of the inner Galaxy, while eliminating the bar's non-axisymmetric resonances and time-dependent torques.

The fifth model, FLHT2024, adds the perturbation from the LMC to FHT2024. To specify the time-dependent LMC perturbation, we first reconstruct an approximate past MW--LMC relative orbit using the framework provided within the \texttt{AGAMA} library\footnote{\url{https://github.com/GalacticDynamics-Oxford/Agama/blob/master/py/example_lmc_mw_interaction.py}}. In this auxiliary orbit calculation, both the MW and the LMC are treated as moving rigid potentials, following a commonly used approximation in recent studies \citep[e.g.,][]{Jethwa2016, Erkal2019, Patel2020, Erkal2021, CorreaMagnus2022, Koposov2023, Brooks2025, Dillamore2026}. 
Their relative orbital history is calculated by integrating backwards in time, accounting for mutual gravitation and dynamical friction acting on the LMC, which is parameterized using the Chandrasekhar formula with a spatially varying Coulomb logarithm. The resulting time-dependent Hamiltonian in the non-inertial Galactocentric frame consists of three components: (1) the static potential of the Milky Way, (2) the moving potential of the LMC, and (3) a time-dependent uniform acceleration field that corrects for the reflex motion of the Galactic center caused by the LMC's pull. While this general approach is common, we note that the specific dynamical friction parameters in the adopted script were originally calibrated for a reference Milky Way potential. Although applying them directly to our modified host potential introduces minor uncertainties, previous studies have demonstrated that such backward-integrated trajectories, employing analytical dynamical friction, generally match the orbital histories derived from full live $N$-body simulations reasonably well \citep[e.g.,][]{Vasiliev2021, CorreaMagnus2022, Lilleengen2023}.

To study the dynamical effects of bar deceleration, we build a final model, FDHT2024, where the Galactic bar slows down following \citet{Chiba2021_sweeping} and \citet{Dillamore2024}. The time evolution of the bar's pattern speed, $\Omega(t)$, is given by:
\begin{equation}
\Omega(t)=
\begin{cases}
\Omega_1[1+\eta\Omega_1 (t+t_0)]^{-1}, & t<t_1,\\[4pt]
\Omega_2\!\left[1+\frac{1}{2}\eta\Omega_2
  (t-t_2)^2/(t_1-t_2)\right]^{-1}, & t_1<t<t_2,\\[8pt]
\Omega_2, & t>t_2,
\end{cases}
\end{equation}
and the transition time $t_2$ is defined as:
\begin{equation}
t_2 = \frac{2}{\eta \Omega_2} - \frac{2}{\eta \Omega_1} - 2t_0 - t_1.
\end{equation}
where we use a constant slowdown rate of $\eta = 0.003$. The initial and final pattern speeds are set to $\Omega_1 = 75.47~\mathrm{km\,s^{-1}\,kpc^{-1}}$ and $\Omega_2 = 34~\mathrm{km\,s^{-1}\,kpc^{-1}}$, respectively, with the time parameters set to $t_0 \approx 3~\mathrm{Gyr}$ and $t_2 \approx 2.7~\mathrm{Gyr}$.

We also adjust the bar length to account for the evolution of the resonance structure. As the bar slows down, its length is expected to increase \citep{1992MNRAS.259..345A}. We therefore scale the bar length dynamically, ensuring that the ratio of the corotation radius to the bar length remains constant. The scaling factor $S$ is given by:
\begin{equation}
S = \frac{\Omega_0}{\Omega(t)},
\end{equation}
where $\Omega_0 = 37.5~\mathrm{km\,s^{-1}\,kpc^{-1}}$ is the reference pattern speed used in the  \citet{hunterTestingKinematicDistances2024} model.

A summary of the specific configurations for each of these six potential models is provided in Table~\ref{tab:potential_models}.


\subsection{Initial conditions} \label{sec:initial_condition}

Because simulating Pal~5 from birth is computationally expensive, we model only the most recent $3~\mathrm{Gyr}$, integrating from $t = -3.0~\mathrm{Gyr}$ to the present day ($t = 0$), which is sufficient to reproduce the observed stream length based on the particle-spray model \citep[e.g.,][]{Bonaca2020}. 

Since Pal~5 is already $\sim11.8$~Gyr old, this setup requires an initial condition that already includes the preceding 8.8~Gyr of stellar and dynamical evolution. We therefore initialize the system at $t=-3$~Gyr using a snapshot at an evolutionary age of $8.8~\mathrm{Gyr}$ from the \texttt{noBin-BH} Pal~5 model of \citet{Wang2024}. That model evolves a Pal~5-like cluster from birth, starting with an isotropic Plummer profile \citep{1911MNRAS..71..460P}, including black hole formation and retention, and excludes primordial binaries to reduce computational cost, as they have negligible effects on the long-term structural evolution and stream morphology \citep{Wang2024}. Dynamically formed binaries during the evolution are still included. The initial stellar masses were sampled from a Kroupa initial mass function \citep[IMF;][]{2001MNRAS.322..231K} in the range $0.1 - 100~M_\odot$, with a metallicity of \(Z=0.0006\) (\([\mathrm{Fe}/\mathrm{H}]\approx -1.4\)) chosen to match observations \citep{Smith2002}.

The \texttt{noBin-BH} model of \citet{Wang2024} includes tidal-stream evolution in the MW2014 potential. To use its snapshots as initial conditions, we first remove tidal-stream stars, allowing us to trace only the stream newly formed over the past 3~Gyr. We apply a ``cut-off radius'', $r_\mathrm{cut}$,  measured from the cluster density center, and retain only stars within this radius. For all models except SHT2024, we adopt $r_\mathrm{cut} = 35~\mathrm{pc}$, corresponding to a tidal radius of $\sim 25~\mathrm{pc}$, resulting in 94,290 objects. Preliminary runs using this filtered snapshot in the SHT2024 potential showed that the stronger tidal field disrupts the cluster too early (Figure~\ref{fig:rh_mass_evolution}). We therefore use a more compact snapshot from an earlier epoch ($8.2~\mathrm{Gyr}$) in the \texttt{noBin-BH} model, with the same $r_\mathrm{cut}$. This yields 102,038 objects and ensures the cluster survives for the full $3~\mathrm{Gyr}$ simulation. Although this introduces a 0.6 Gyr age offset, the impact on stellar evolution is expected to be small at this late stage of evolution. Table~\ref{tab:initial_conditions} summarizes the initial conditions for our simulations.


\begin{deluxetable}{lccc}
\tablecaption{Initial Cluster Properties at $t = -3~\mathrm{Gyr}$\label{tab:initial_conditions}}
\tablehead{
\colhead{Parameter} & \colhead{Unit} & \colhead{Standard Models} & \colhead{SHT2024 Model}
}
\startdata
Source Snapshot (Age)       & (Gyr)   & 8.8 (noBin-BH) & 8.2 (noBin-BH) \\
Cut-off radius ($r_{\rm cut}$) & (pc)    & 35             & 35             \\
Number of particles ($N$)   & ---     & 94,290         & 102,038         \\
Half-mass radius ($r_h$)    & (pc)    & 16.17          & 16.20          \\
Total initial mass ($M_{\rm ini}$) & ($M_\odot$) & 38,287.1       & 41,135.5       \\
\enddata
\tablecomments{All initial models are dynamically evolved snapshots extracted from the parent direct $N$-body simulations of \citet{Wang2024}. The properties listed here are measured at the start of our orbital integration ($t = -3.0~\mathrm{Gyr}$), after filtering stars outside the specified cut-off radius ($r_{\rm cut}$). The total initial mass ($M_{\rm ini}$) is computed as twice the bound mass within $r_h$.}
\end{deluxetable}


\begin{deluxetable}{lcc}
\tablecaption{Parameters used for Pal~5\label{tab:pal5_params}}
\tablehead{
\colhead{Parameter} & \colhead{Unit} & \colhead{Value}
}
\startdata
RA (J2000) & (deg) & 229.0217 \\
Dec (J2000) & (deg) & $-0.1109$ \\
Radial velocity & (km\,s$^{-1}$) & $-57.5$ \\
PM$_{\rm RA}\cos({\rm Dec})$ & (mas\,yr$^{-1}$) & $-2.67$ \\
PM$_{\rm Dec}$ & (mas\,yr$^{-1}$) & $-2.67$ \\
Distance from Sun & (kpc) & 19.98 \\
\enddata
\end{deluxetable}


We determine the present-day Cartesian Galactocentric position and velocity of Pal~5 through a coordinate transformation process based on the best-fit model of \citet{Gieles2021}. First, we recover the observable quantities (RA, Dec, distance, proper motions, and radial velocity) from their reported Cartesian coordinates, using the Solar parameters adopted in their work (Solar position $[-8.182, 0, 0]~\mathrm{kpc}$ and velocity $[11.1, 245.7, 7.3]~\mathrm{km\,s^{-1}}$). The resulting observational parameters are listed in Table~\ref{tab:pal5_params}. Next, we transform these observables back into Galactocentric Cartesian coordinates using our updated reference frame, where the Sun is placed at $[-8.178, 0, 0.021]~\mathrm{kpc}$ with a velocity of $[11.1, 241.24, 7.25]~\mathrm{km\,s^{-1}}$ \citep{Gravity2019,SBD10}. The final adopted coordinates are $[x, y, z] = [5.771, 0.2069, 14.32]~\mathrm{kpc}$ and $[v_x, v_y, v_z] = [-41.36, -116.37, -16.74]~\mathrm{km\,s^{-1}}$. We note that these values differ slightly from those in \citet{Gieles2021} (e.g., $x=5.733~\mathrm{kpc}$ in their fit), strictly due to the difference in the adopted Solar position and velocity. We do not explore the observational uncertainty space (e.g., via MCMC sampling), as fixing the present-day coordinates provides a strict control variable for comparing macroscopic kinematic effects across different potentials. Furthermore, while the true forward-integrated orbit in a fully self-gravitating $N$-body simulation naturally deviates slightly from a pure test-particle backward trajectory due to asymmetric mass loss in a time-dependent potential, iteratively fine-tuning the initial coordinates to perfectly compensate for this effect across multiple time-dependent potentials remains computationally prohibitive.

\section{Results} \label{sec:results}


\subsection{Orbits in the Galaxy}

\begin{deluxetable}{lcccccc}
\tablecaption{Initial positions and velocities for different Galactic potential models\label{tab:init_pos_vel}}
\tablehead{
\colhead{Model} &
\colhead{$x_i$ (kpc)} & \colhead{$y_i$ (kpc)} & \colhead{$z_i$ (kpc)} &
\colhead{$v_{x,i}$ (km\,s$^{-1}$)} & \colhead{$v_{y,i}$ (km\,s$^{-1}$)} & \colhead{$v_{z,i}$ (km\,s$^{-1}$)}
}
\startdata
MW2014   & $-5.143$ &  4.031 & $-7.522$ &  218.0 & $-42.3$ & $-27.2$ \\
SHT2024  & 13.447   & $-4.048$ & $-4.308$ & $-17.6$ & $-43.7$ &  143.8 \\
FHT2024  & $-7.825$ & 10.972 &  5.281 & $-11.4$ &  96.4 & $-122.4$ \\
FNHT2024 & $-7.602$ & 13.463 &  0.820 &  36.2 &  21.6 & $-145.1$ \\
FLHT2024 & $-7.817$ &  9.843 &  7.153 & $-19.1$ & 108.9 & $-104.9$ \\
FDHT2024 & $-8.077$ & 12.387 & $-2.160$ &  79.7 & $-40.5$ & $-135.7$ \\
\enddata
\end{deluxetable}


\begin{figure}[htbp]
    \centering
    \includegraphics[width=\textwidth]{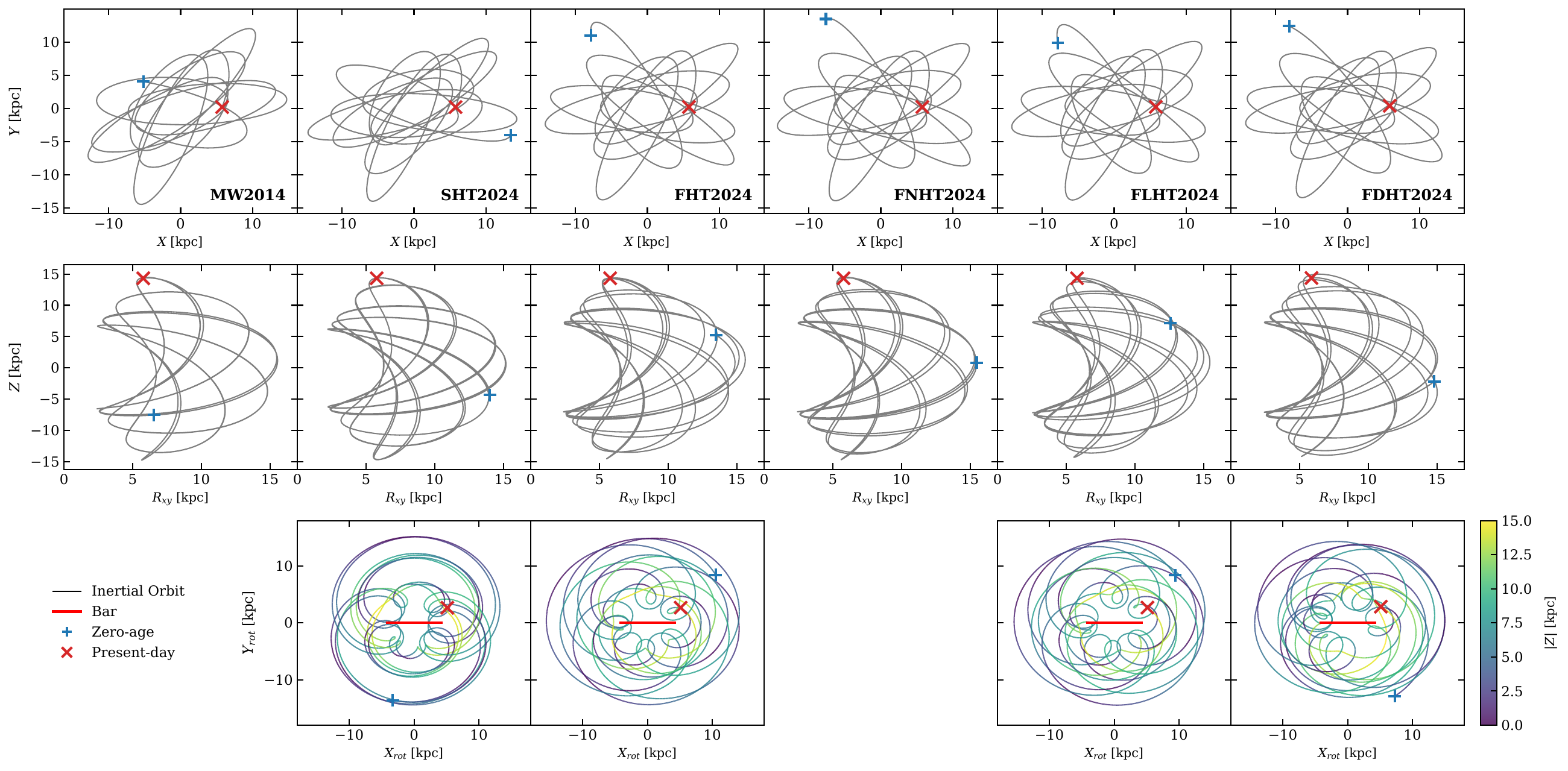}
    \caption{
        Orbital evolution of the Pal~5 cluster across six potential models (columns).
        The top row displays the trajectory in the inertial Galactocentric $X$--$Y$ frame.
        The middle row shows the projection in the meridional plane ($R_\text{G}$--$Z$), where $R_\text{G} = \sqrt{X^2 + Y^2}$.
        The bottom row presents the orbit in the bar-corotating frame ($X_{rot}$--$Y_{rot}$), where the color scale indicates the absolute vertical distance $|Z|$ from the Galactic midplane.
        In the bottom panels, the solid red line indicates the orientation of the Galactic bar along the $X_{rot}$-axis.
        Note that the first column (MW2014) represents an axisymmetric potential without a bar component; therefore, the corresponding corotating frame is not shown.
        The symbols `+' and `$\times$' mark the initial position at $t = -3~\mathrm{Gyr}$ and the present-day position at $t = 0$, respectively.
    }
    \label{fig:pal5_orbit_comparison}
\end{figure}


\begin{figure}[htbp]
    \centering
    \includegraphics[width=\textwidth]{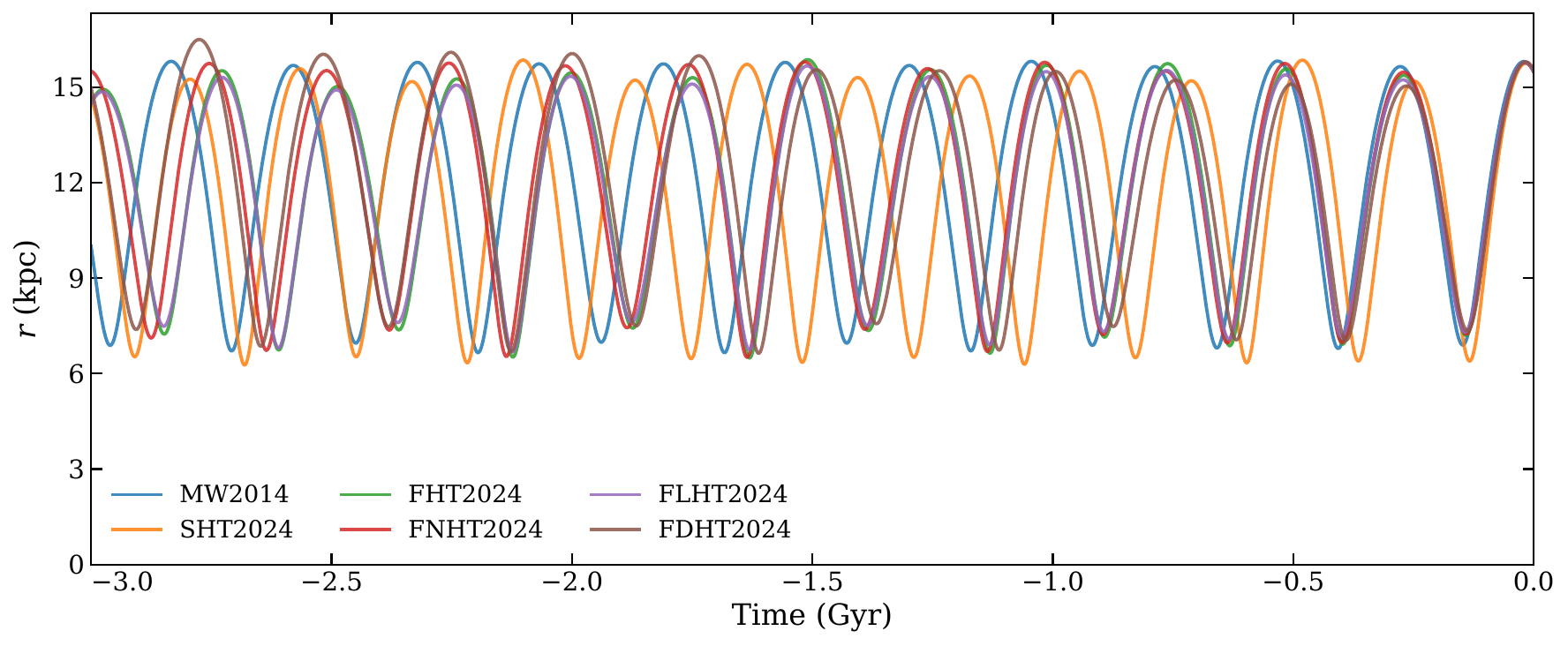}
    \caption{
        Time evolution of the Galactocentric distance ($r$) for the Pal~5 progenitor across the six potential models over the full $3~\mathrm{Gyr}$ backward integration (from $t=-3~\mathrm{Gyr}$ to the present day $t=0$).
    }
    \label{fig:orbit_r_t}
\end{figure}


\begin{deluxetable}{lcccc}
\tablecaption{Orbital Properties of the Pal~5 Progenitor Across Different Potentials \label{tab:orbital_properties}}
\tablehead{
\colhead{Potential Model} & \colhead{$r_{\rm peri}$} & \colhead{$r_{\rm apo}$} & \colhead{Eccentricity} & \colhead{$N_{\rm peri}$} \\
\colhead{} & \colhead{(kpc)} & \colhead{(kpc)} & \colhead{($e$)} & \colhead{}
}
\startdata
MW2014     & 6.65 & 15.81 & 0.408 & 12 \\
SHT2024    & 6.27 & 15.85 & 0.433 & 13 \\
FHT2024    & 6.49 & 15.86 & 0.419 & 12 \\
FNHT2024   & 6.51 & 15.79 & 0.416 & 12 \\
FLHT2024   & 6.68 & 15.77 & 0.405 & 12 \\
FDHT2024   & 6.64 & 16.49 & 0.426 & 12 \\
\enddata
\tablecomments{Orbital properties are derived from the $3~\mathrm{Gyr}$ backward integration. $r_{\rm peri}$ and $r_{\rm apo}$ represent the mean pericentric and apocentric distances, respectively. The eccentricity is calculated as $e = (r_{\rm apo} - r_{\rm peri})/(r_{\rm apo} + r_{\rm peri})$. $N_{\rm peri}$ denotes the total number of pericentric passages during the $3~\mathrm{Gyr}$ period.}
\end{deluxetable}


To find the initial position and velocity of the Pal~5 progenitor in the Galaxy, we rewind its orbit by integrating backward in time for $3~\mathrm{Gyr}$ across all potential models, treating the cluster as a point mass. Since the gravitational potential is different in each model (as described in Section~\ref{sec:potential}), this backward integration gives different initial positions and velocities for each case, as listed in Table~\ref{tab:init_pos_vel}. The resulting forward-evolved trajectories of the surviving Pal~5 core are displayed in Figure~\ref{fig:pal5_orbit_comparison}. In each simulation snapshot, the core's position is determined by identifying the density peak (density center) of the bound particles \citep{Casertano1985}.

In the inertial Galactocentric $X$--$Y$ frame (top row), the trajectories of the MW2014 and SHT2024 models appear relatively similar, both exhibiting slow precession over the 3\,Gyr integration. In contrast, the remaining four models (FHT2024, FNHT2024, FLHT2024, and FDHT2024) show a distinct behavior where the flattening of the dark matter halo significantly increases the orbital precession rate. This causes the orbital loops to cover a much wider azimuthal range within the same duration compared to the MW2014 and SHT2024 case.

In the meridional $R_{xy}$--$Z$ plane (middle row), the MW2014 model exhibits a highly regular, Lissajous-like orbital pattern, reflecting a stable trajectory within a spherical dark matter halo. Visually, the introduction of the Galactic bar in the SHT2024 model produces only minor deviations in the macroscopic center-of-mass trajectory; the overall orbital envelope remains largely similar to that of MW2014 over this $3~\mathrm{Gyr}$ integration. The macroscopic orbital morphology shifts much more significantly with the introduction of a flattened dark matter halo (e.g., FHT2024 and subsequent models). The flattened potential strongly alters the vertical and radial restoring forces, causing a more pronounced orbital precession and noticeably shifting the turning points compared to the tightly interlaced regular tracks seen in the spherical halo models.

In the bar-corotating frame (the bottom row of Figure~\ref{fig:pal5_orbit_comparison}), the SHT2024 model displays a highly regular, ``peanut''-shaped trajectory with repetitive closed loops. Such morphology strongly indicates a resonant capture. To rigorously verify this, we conducted a frequency analysis of the cluster's orbit. We find that the azimuthal and radial frequencies ($\Omega_\phi \approx -18.7~\mathrm{rad\,Gyr^{-1}}$, $\Omega_R \approx 38.8~\mathrm{rad\,Gyr^{-1}}$) exhibit a strict low-order commensurability with the bar's pattern speed ($\Omega_{\rm bar} \approx -38.4~\mathrm{rad\,Gyr^{-1}}$), yielding a frequency ratio of $|(\Omega_\phi - \Omega_{\rm bar}) / \Omega_R| \approx 0.506$. The libration of the corresponding resonant angle confirms that the Pal~5 progenitor is trapped in a 2:1 resonance (specifically, the Outer Lindblad Resonance) with the Galactic bar in the spherical halo potential (see Appendix~\ref{sec:appendix_resonance} for the libration plot).

However, the introduction of the flattened dark matter halo alters the vertical and radial restoring forces, breaking this delicate resonant phase-locking. Consequently, in the FHT2024 and FLHT2024 models, the orbits no longer follow repetitive resonant paths but instead exhibit prominent secular precession. The FDHT2024 model exhibits a significantly more complex and disordered trajectory, which is primarily induced by the time-dependent deceleration of the Galactic bar sweeping through resonance parameter space.

Figure~\ref{fig:orbit_r_t} illustrates the time evolution of the Galactocentric distance $r$ over the full $3~\mathrm{Gyr}$ backward integration. To quantitatively compare the orbital properties driving the disruption of the cluster, we extract the minimum pericentric distance ($r_{\rm peri}$), maximum apocentric distance ($r_{\rm apo}$), orbital eccentricity ($e$), and the total number of pericentric passages ($N_{\rm peri}$) for each model, as summarized in Table~\ref{tab:orbital_properties}. A direct comparison reveals that the SHT2024 model penetrates deeper into the Galactic center, reaching a smaller pericentric distance ($r_{\rm peri} = 6.27~\mathrm{kpc}$) and exhibiting a higher eccentricity ($e = 0.433$) than the baseline MW2014 model ($r_{\rm peri} = 6.65~\mathrm{kpc}$, $e = 0.408$). Furthermore, the orbital frequency in the SHT2024 potential is notably higher, resulting in one additional pericentric passage ($N_{\rm peri} = 13$) compared to all other models ($N_{\rm peri} = 12$) over the $3~\mathrm{Gyr}$ period. These combined physical factors, closer pericentric encounters and an increased frequency of tidal shocks, directly account for the accelerated mass loss and early disruption observed in the SHT2024 model.

The final center-of-mass position of the simulated cluster after 3~Gyr of forward evolution does not perfectly coincide with the initial observed position used for the backward integration. As shown in Table~\ref{tab:final_positions}, there are small spatial offsets ($\Delta r$) ranging from $\sim 0.001$ to $0.089~\mathrm{kpc}$ among the models. This drift occurs because the backward integration assumes a rigid point mass, whereas the forward simulation models an extended $N$-body system subject to asymmetric mass loss and internal dynamical evolution. To account for this, whenever a comparison with observational data is required in the subsequent analysis, we correct the positions of the stars by applying a spatial offset to align the center-of-mass of the simulated $N$-body models with the observed position of Pal~5.


\begin{table}
	\centering
	\caption{Final center-of-mass positions of the simulated clusters after 3~Gyr of evolution and their spatial offsets ($\Delta r$) from the observed position of Pal~5. \label{tab:final_positions}}
	\begin{tabular}{lccccc}
		\hline
		Model & $x_f$ & $y_f$ & $z_f$ & $\Delta r$ \\
		& (kpc) & (kpc) & (kpc) & (kpc) \\
		\hline
		MW2014   & 5.772 & 0.206 & 14.325 & 0.001 \\
		SHT2024  & 5.772 & 0.206 & 14.325 & 0.002 \\
		FHT2024  & 5.779 & 0.225 & 14.329 & 0.020 \\
        FNHT2024  & 5.779 & 0.219 & 14.328 & 0.015 \\
		FLHT2024 & 5.783 & 0.238 & 14.329 & 0.034 \\
		FDHT2024 & 5.741 & 0.119 & 14.312 & 0.089 \\
		\hline
	\end{tabular}
\end{table}

\subsection{Cluster Evolution}


\begin{figure}[htbp]
    \centering
    \includegraphics[width=\textwidth]{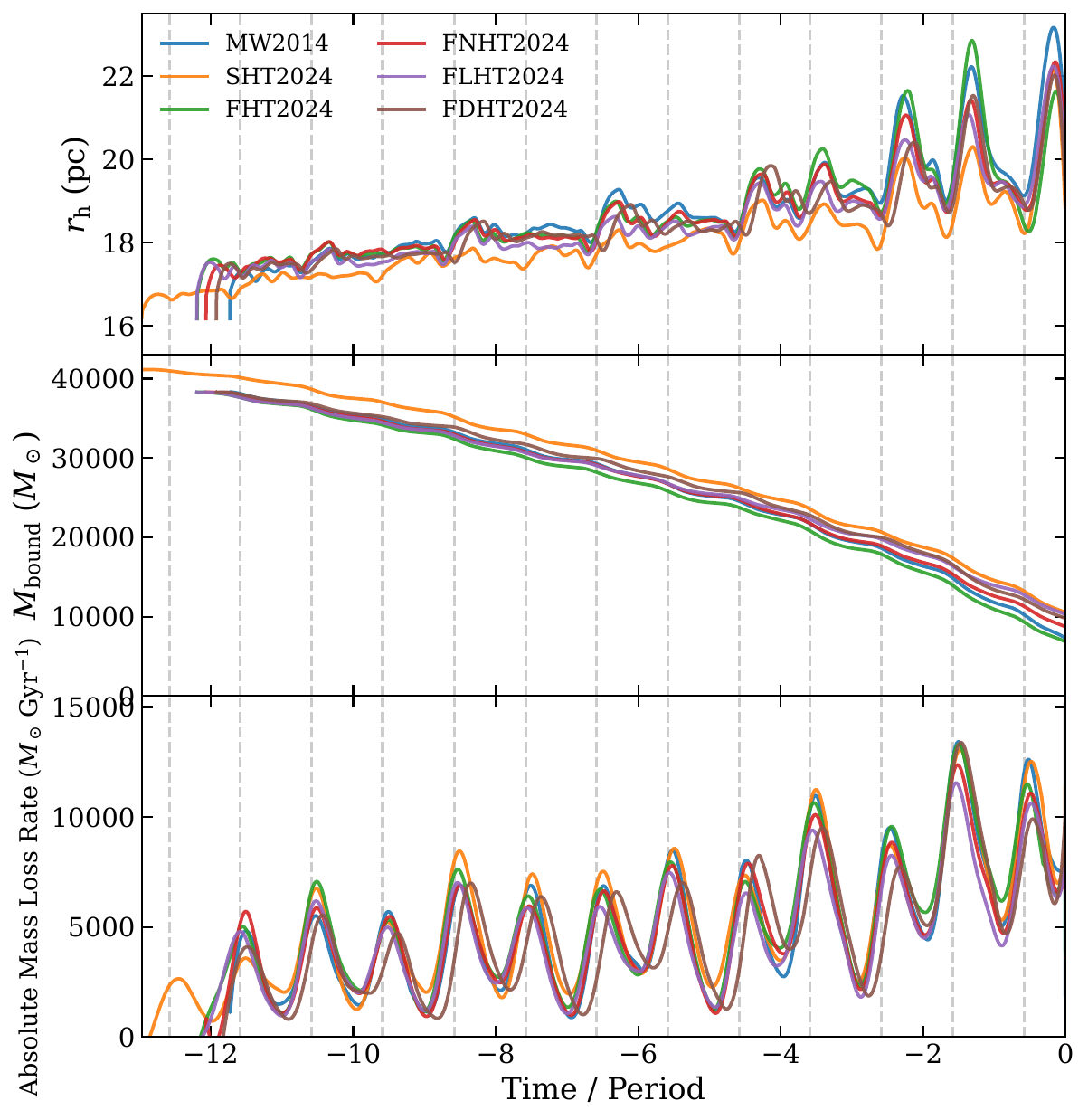}
    \caption{
        Structural evolution and absolute mass-loss rate of the Pal~5 progenitor across the tested Galactic potentials.
        To eliminate apparent phase shifts caused by varying orbital frequencies, the horizontal axis is normalized by the average orbital period ($P$) of each model, ending at the present day ($0$). 
        The upper panel displays the smoothed half-mass radius ($r_{\mathrm{h}}$), the middle panel shows the bound mass evolution ($M_{\mathrm{bound}}$), while the lower panel shows the corresponding absolute mass-loss rate ($dM/dt$).
        Vertical dashed lines indicate the true orbital pericentric passages for the SHT2024 model. 
        Note that the SHT2024 model requires a more compact initial structure and a larger total mass to ensure the survival of its core over the full $3~\mathrm{Gyr}$ integration (see Section~\ref{sec:initial_condition} for details).
    }
    \label{fig:rh_mass_evolution}
\end{figure}


\begin{table}[htbp]
\centering
\caption{Initial and Final Mass Properties of the Pal~5 Progenitor Across Different Potentials}
\label{tab:mass_summary}
\begin{tabular}{lccc}
\hline
\hline
Potential Model & Initial Mass & Final Mass & Remaining Fraction \\
 & ($M_\odot$) & ($M_\odot$) & (\%) \\
\hline
MW2014   & 38,287.1 & 7,357.1  & 19.2 \\
SHT2024  & 41,135.5 & 10,530.5  & 25.6 \\
FHT2024  & 38,287.1 & 6,843.6  & 17.9 \\
FNHT2024 & 38,287.1 & 8,729.1  & 22.8 \\
FLHT2024 & 38,287.1 & 10,297.0 & 26.9 \\
FDHT2024 & 38,287.1 & 9,746.9  & 25.5 \\
\hline
\end{tabular}
\end{table}


\subsubsection{Mass Loss} \label{sec:mass_loss}


The mass loss of the cluster happens through two main ways: wind mass loss driven by stellar evolution and the escape of stars caused by the cluster's internal dynamics and external tidal stripping. The latter has dominated the past 3~Gyr of Pal~5 evolution. To define the bound mass consistently across all simulations, all models use the same criteria for identifying escaping stars. Specifically, a particle is considered an escaper only if it meets both of the following conditions: (1) its distance from the cluster center is larger than twenty times the current half-mass radius ($r > 20~ r_{\mathrm{h}}$); and (2) its total specific energy—calculated in isolation without the external Galactic potential—is positive ($E > 0$).

Figure~\ref{fig:rh_mass_evolution} presents the structural evolution and the absolute mass-loss rate of the Pal~5 progenitor across the tested Galactic potentials. To eliminate apparent phase shifts caused by varying orbital frequencies and to directly align the orbital phases, the horizontal axis is normalized by the average orbital period ($P$) for each model. The top panel displays the evolution of the smoothed half-mass radius ($r_{\mathrm{h}}$). In general, all models exhibit a secular expansion in $r_{\mathrm{h}}$ over time, driven by internal two-body relaxation and continuous tidal stripping. Superimposed on this overall expansion trend are sharp, periodic structural compressions corresponding to the pericentric passages. The bottom panel illustrates the corresponding absolute mass-loss rate ($dM/dt$), which exhibits distinct periodic peaks, explicitly demonstrating that the mass depletion is dominantly driven by these tidal shocks.

To ensure its survival over the full integration, the SHT2024 model was deliberately initialized with a significantly more massive and compact structure. Structurally, this deeper internal potential well tightly binds the core, causing the SHT2024 model to maintain a systematically smaller half-mass radius ($r_{\mathrm{h}}$) throughout the simulation (Figure~\ref{fig:rh_mass_evolution}, upper panel). This extreme initial compactness also acts as a highly effective shield during the earliest phase of its evolution. As shown in the bottom panel, the stronger self-gravity allows the SHT2024 model to successfully resist the initial tidal perturbations, exhibiting comparatively low absolute mass-loss rates prior to $t/P \approx -11$.

However, due to its smaller average pericentric distance ($r_{\mathrm{peri}} = 6.27~\mathrm{kpc}$, compared to a minimum of $6.49~\mathrm{kpc}$ for the other models; see Table~\ref{tab:orbital_properties}), the SHT2024 model inevitably experiences a higher absolute mass-loss rate after $t/P \approx -11$. This closer approach to the Galactic center subjects the progenitor to a substantially stronger tidal field during each pericentric passage, easily overwhelming its initial structural advantage. Consequently, the SHT2024 model suffers severe, periodic mass-loss peaks. Combined with its higher orbital frequency, which contributes one additional pericentric stripping event over the $3~\mathrm{Gyr}$ period, this extreme tidal environment fundamentally dominates the long-term dynamical evolution, ultimately leading to a substantial absolute mass loss.

The FLHT2024 model, which incorporates the gravitational influence of the LMC, successfully preserves the largest fraction of its initial mass ($26.9\%$) and ends the simulation with the highest final bound mass among the standard-mass models ($10,297.00~M_\odot$). Consistent with this reduced disruption efficiency, the bottom panel of Figure~\ref{fig:rh_mass_evolution} shows that the FLHT2024 model systematically exhibits the lowest peaks in the fractional mass-loss rate throughout the entire integration. Driven by the reflex motion of the Milky Way barycenter in response to the massive LMC infall \citep[e.g.,][]{Erkal2019, Petersen2020, Vasiliev2021}, the average pericentric distance of the cluster is lifted to $6.68~\mathrm{kpc}$ (compared to $6.49~\mathrm{kpc}$ in the baseline FHT2024 model), representing the shallowest pericentric plunge among all tested potentials (Table~\ref{tab:orbital_properties}). Because the efficiency of tidal stripping is highly sensitive to pericentric depth, this $\sim 0.19~\mathrm{kpc}$ increase systematically mitigates the intensity of the compressive tidal shocks during each of the 12 pericentric passages. The continuous accumulation of these mitigated shocks over the $3~\mathrm{Gyr}$ dynamical evolution ultimately results in the most gentle mass-loss history for the FLHT2024 model.

The remaining models focus on the influence of the Galactic bar. By comparing the FHT2024 (constant-speed bar), FNHT2024 (no bar), and FDHT2024 (decelerating bar) models, it becomes evident that while the macroscopic shapes of their orbital tracks remain largely similar, the different treatments of the central bar component lead to subtle but consequential differences in the average pericentric distance. Specifically, these distinct bar configurations yield average pericentric depths of $6.49~\mathrm{kpc}$ for FHT2024, $6.51~\mathrm{kpc}$ for FNHT2024, and $6.64~\mathrm{kpc}$ for FDHT2024 (Table~\ref{tab:orbital_properties}). As established previously, the efficiency of tidal stripping is extremely sensitive to this pericentric depth. Consequently, the models that maintain slightly larger pericentric distances effectively mitigate the intensity of compressive tidal shocks, thereby preserving a higher fraction of their initial mass ($22.8\%$ for FNHT2024 and $25.5\%$ for FDHT2024) compared to the deeper pericentric plunge of the baseline FHT2024 model ($17.9\%$). Ultimately, these cases demonstrate that the Galactic bar dynamically modulates the mass-depletion history of Pal~5 by changing its pericentric evolution.


\begin{figure}[htbp]
\centering
\includegraphics[width=\textwidth]{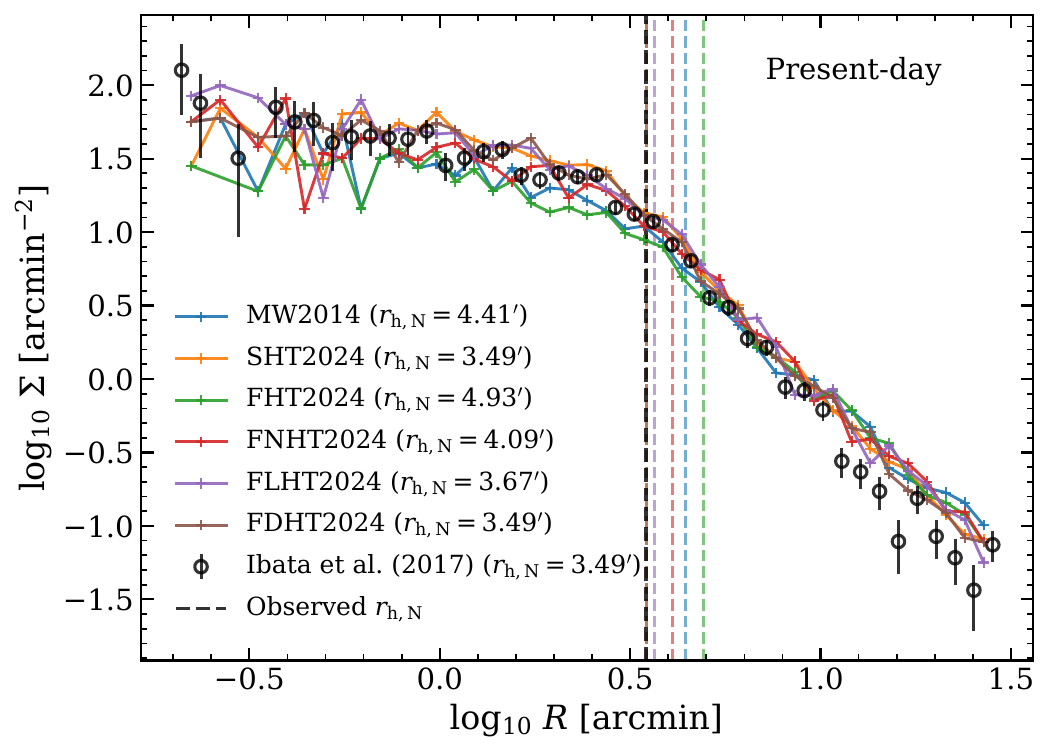}
\caption{
The surface number density $\Sigma(R)$ profiles are shown for the $N$-body models along with observational data from \citet{Ibata2017-Pal5}.
The colored curves represent the simulation results at the present day for different Galactic potential models.
Black circles show the observed profile, with error bars indicating uncertainties.
Vertical lines are used to indicate the ``effective radius'' , the radius containing half the number of stars in projection, ($R_{\mathrm{h,N}}$) of the clusters.
}
\label{fig:profile}
\end{figure}


\subsubsection{Surface number density profiles} \label{sec:profiles}


To validate the structural properties of the surviving cluster, Figure~\ref{fig:profile} compares the projected surface number density ($\Sigma$) profiles of our models at the end of the integration with the observed profile of Pal~5 \citep{Ibata2017-Pal5}. To strictly match the observational selection function, we restrict the simulated $N$-body data to main-sequence stars with masses between $0.625\,M_\odot$ and $0.815\,M_\odot$ \citep[see][for details]{Gieles2021}, and compute the density profiles as a function of the angular distance $R$ from the cluster center.

Figure~\ref{fig:profile} presents the projected surface number density ($\Sigma$) profiles for our simulated models alongside the observational data. In the outer regions, the differences among the various models are relatively small. However, beyond $\log_{10}(R) \approx 1.0$, all simulations systematically overpredict the surface density compared to the observed profile. This discrepancy may partly reflect observational challenges in the low-density outskirts, where unambiguously identifying loosely bound cluster members against the field background becomes increasingly difficult \citep[e.g.,][]{Odenkirchen_2003, Ibata2017-Pal5}. A similar tendency for the simulated profile to lie above the observed profile in the outskirts is also visible in the comparison presented by \citet{Gieles2021}. Because this observational limitation inevitably misses extended stellar populations, the observationally derived half-number radius ($R_{\mathrm{h,N}}$) is artificially biased toward a smaller value, effectively acting as an observational lower limit. Consequently, the simulated $R_{\mathrm{h,N}}$ values (indicated by the vertical dashed lines), which fully account for the extended structures, are naturally expected to be larger than or equal to this biased observational benchmark.

In contrast to the outskirts, the surface density in the inner regions exhibits noticeable variations among the models, which are correlated with their final remaining bound masses. The SHT2024, FLHT2024, and FDHT2024 models, which retain relatively higher masses (approximately $10,000~M_\odot$), share similar spatial distributions. Their remaining masses allow them to maintain central regions that trace the inner observational data reasonably well. Despite preserving more stars in their extended outskirts, the relatively concentrated cores of the SHT2024 and FDHT2024 models influence the overall spatial distribution, yielding half-number radii of $3.49^\prime$. This indicates a structural concentration comparable to the observational limit, and the FLHT2024 model ($3.67^\prime$) is also consistent with this lower bound. Conversely, models that experience more mass loss, namely the MW2014 and baseline FHT2024 models (retaining approximately $7,000~M_\odot$), exhibit lower inner surface densities. With fewer stars in their central regions due to tidal stripping, the surviving populations in these models form a relatively more diffuse spatial distribution, leading to larger half-number radii ($4.41^\prime$ and $4.93^\prime$, respectively).

The simulated profiles are comparable to the observational data in order of magnitude. This confirms that the used initial cluster parameters are reasonable and that the cluster does not completely dissolve by the end of the simulation.


\subsection{Tidal stream}
\label{sec:tidal_stream}


\subsubsection{Construction of the observational stream measurements}
\label{sec:obs_stream_measurements}


\begin{figure}[htbp]
    \centering
    \includegraphics[width=\textwidth]{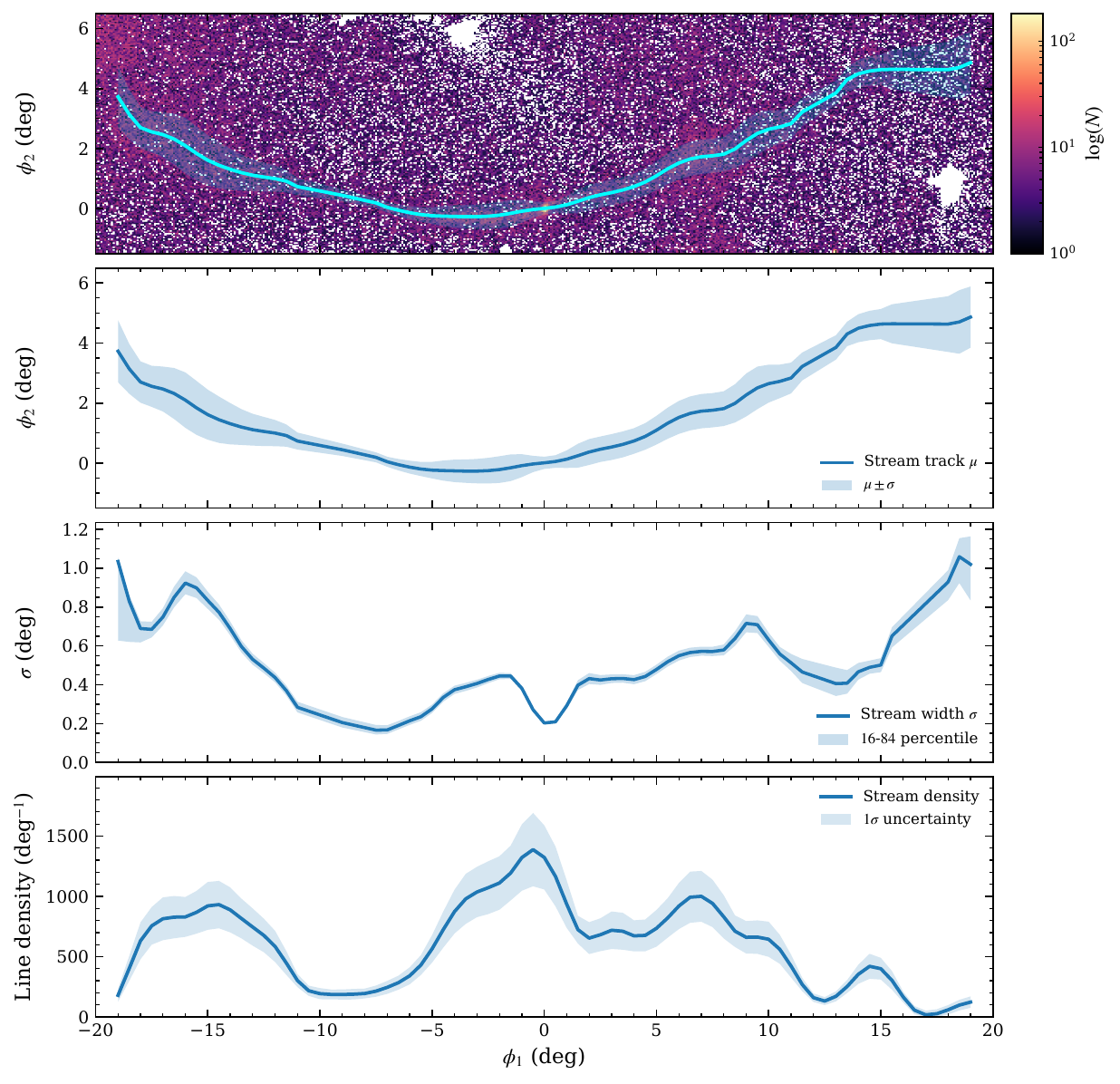}
    \caption{
    Visualization of the Palomar~5 tidal stream in the Pal~5 stream coordinate
    system implemented in \texttt{gala} as \texttt{Pal5PriceWhelan18}.
    The top panel shows the spatial distribution of candidate stream stars in the
    $(\phi_1,\phi_2)$ plane, with the stellar density color-coded on a logarithmic
    scale. The blue curve marks the best-fit stream track, $\mu(\phi_1)$, and the
    shaded band indicates the corresponding $\mu \pm \sigma$ region.
    The second panel shows the stream track $\mu(\phi_1)$ inferred from the
    guided MCMC fitting procedure, with the shaded region representing the
    transverse stream width, $\mu \pm \sigma$.
    The third panel shows the stream width $\sigma(\phi_1)$ as a function of
    $\phi_1$, with the shaded region indicating the 16th--84th percentile
    confidence interval.
    The bottom panel presents the background-subtracted line-density profile
    along the stream, with the shaded band denoting the $1\sigma$ uncertainty.
    }
    \label{fig:pal5_spatial_morphology_track_width_density}
\end{figure}

A direct comparison between the simulated Palomar~5 streams and observations
requires an observational data set that traces both tidal tails over a large angular
extent. We therefore use the DESI Legacy Imaging Surveys DR10 data set analysed by
\citet{Xiao2025}, which extends the detected leading tail to approximately
\(\delta\simeq -15^\circ\), as the basis for constructing the observational
reference profiles of the stream morphology, track, width, and density. Through
communication with the authors, we obtained the pre-processed DR10 photometric
catalogue used in their analysis.

Starting from this catalogue, we follow the colour--colour selection, HDBSCAN
clustering, and membership-assignment procedure of \citet{Xiao2025}. Since the exact
isochrone implementation used for their CMD filtering is not fully specified in
\citet{Xiao2025}, we independently construct a PARSEC isochrone to reproduce the CMD
selection. Specifically, we generate the isochrone using the PARSEC v1.2S models
through the CMD web interface, adopting the DECam AB-magnitude photometric system and
OBC bolometric corrections. The selected isochrone has a metal mass fraction
\(Z=2.47\times10^{-4}\), corresponding to a global metallicity
\([{\rm M/H}]=-1.798\), and \(\log_{10}({\rm age/yr})=10.061\), corresponding to an
age of approximately \(11.5\) Gyr. The absolute magnitudes in the \texttt{DES-gmag},
\texttt{DES-rmag}, and \texttt{DES-zmag} bands are shifted by the Palomar~5 distance
modulus, \(DM=16.835\), before applying the CMD selection.

The resulting photometrically selected sample is then processed with the same
HDBSCAN-based member-selection strategy as \citet{Xiao2025}, including the
six-dimensional photometric feature space \((g,r,z,g-r,g-z,r-z)\), the adopted
clustering parameters, and the final membership-distance threshold. These selected
candidates are used to construct the observational stream track, transverse width,
and line-density profile for comparison with our \(N\)-body models.

Using this selected candidate sample, we transform the stars into the
\texttt{Pal5PriceWhelan18} stream coordinate frame implemented in \texttt{gala}
\citep{PriceWhelan2017}, and measure the stream profiles over
\(-20^\circ<\phi_1<20^\circ\). The stream track and width are measured in overlapping
\(\phi_1\) bins with a bin width of \(2^\circ\) and a step size of \(0.5^\circ\).
In each bin, the transverse \(\phi_2\) distribution is modeled as a Gaussian stream
component plus a locally linear background, following the same general form as
\citet{Xiao2025}. The fitted parameters are the stream fraction \(f\), centroid
\(\mu\), width \(\sigma\), and the slope and intercept of the linear background.
To stabilize the fit in low-density regions, the MCMC initialization is guided by a
smooth ridge estimate of the two-dimensional density map. We use the median posterior
values of \(\mu\) and \(\sigma\) as the stream track and width, respectively, and
adopt the 16th--84th percentile range of the posterior samples as the corresponding
\(1\sigma\)-equivalent credible interval. We also construct a background-subtracted
line-density profile by counting stars in an on-stream aperture centred on the fitted
track and subtracting local off-stream sidebands. These track, width, and line-density
measurements are the Xiao+2025-based observational profiles used in the comparisons
below.

Figure~\ref{fig:pal5_spatial_morphology_track_width_density} summarizes the resulting
observational measurements. The upper panel shows the two-dimensional density map of
the selected candidate stars in the \texttt{gala} Palomar~5 stream coordinates, with
the fitted track and \(\mu\pm\sigma\) band overlaid. The following panels show the
track, width, and background-subtracted line density. Overall, our reconstruction
recovers the same large-scale Palomar~5 stream morphology, track, width, and density
behaviour reported by \citet{Xiao2025}.

Our reconstructed profile is qualitatively consistent with the large-scale density
structure reported by \citet{Bonaca2020}. In particular, \citet{Bonaca2020}
identified two prominent underdensities in the Palomar~5 stream at approximately
\(\phi_1\simeq -7^\circ\) and \(\phi_1\simeq 3^\circ\). Similar low-density
features are present in our reconstructed line-density profile, although their
detailed amplitudes depend on the adopted member selection and background
subtraction. However, we do not find clear evidence for the strong fan-like
broadening near \(\phi_1\sim5^\circ\) reported by \citet{Bonaca2020}. For the
quantitative comparisons below, we use the publicly tabulated line-density and width
measurements of \citet{Erkal2017}, together with our reconstructed Xiao+2025
measurements.


\subsubsection{Stream morphology, track, and width}
\label{sec:stream_morphology_width}


\begin{figure*}[htbp]
\centering
\includegraphics[width=\textwidth]{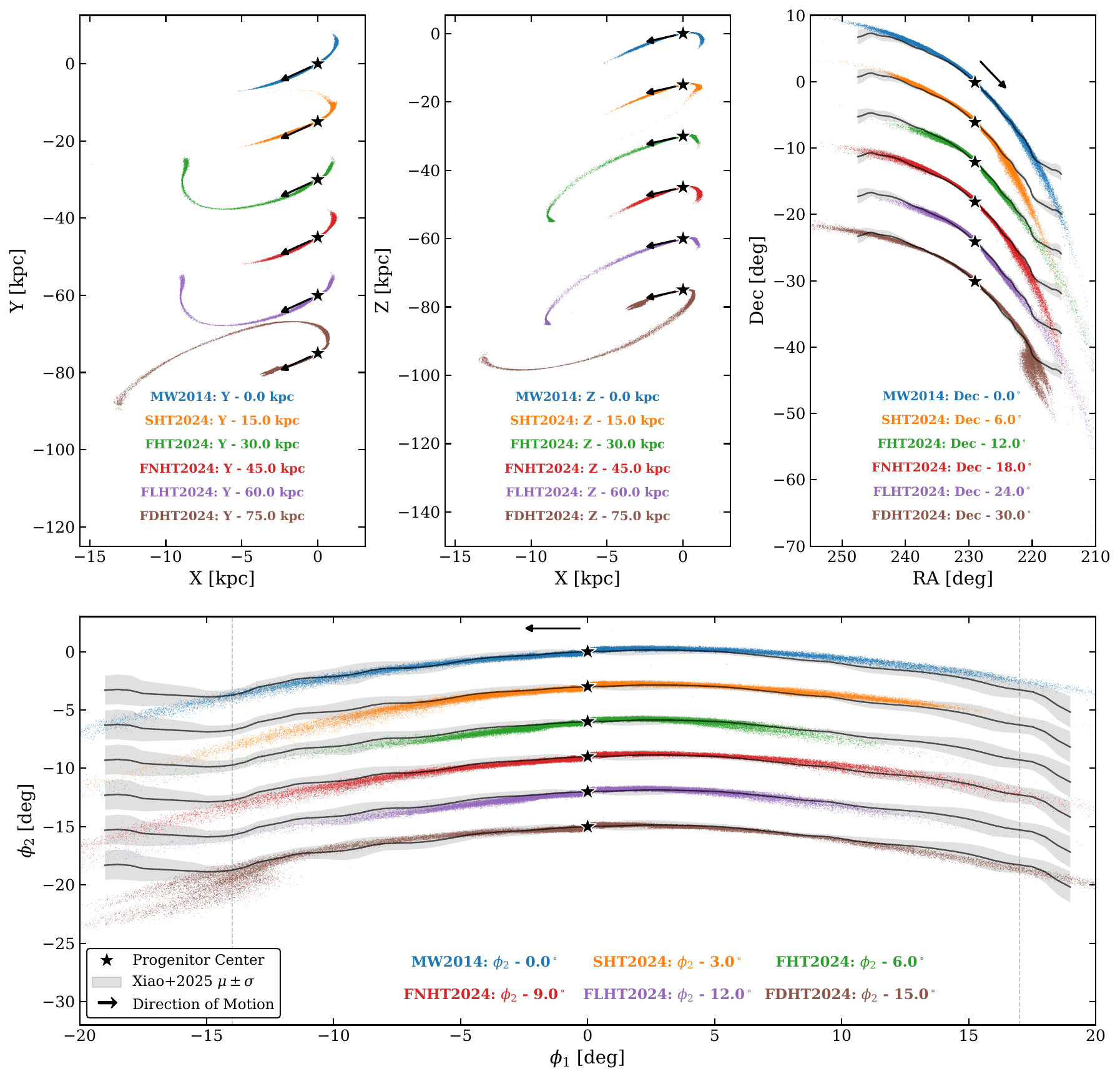}
\caption{
Spatial morphology of the simulated Palomar~5 streams across the tested Galactic potentials at the present day.
The top row displays the stream distributions in the Galactocentric Cartesian $X$--$Y$ and $X$--$Z$ planes, followed by the equatorial projection in RA--Dec coordinates.
The bottom row presents the on-sky morphology in the rotated stream coordinate system $(\phi_1, \phi_2)$, where the coordinate transformation is adopted from \citet{Erkal2017}.
To facilitate a clear visual comparison of the stream lengths and morphologies, the simulated particles for each model are artificially shifted vertically. The coordinate shifts applied to each model (e.g., $Y - \Delta Y$) are explicitly labeled below each track to indicate how the plotted positions are derived.
The physical location of the Pal~5 progenitor is marked by a black star in all panels, corresponding to the theoretical origin $(\phi_1, \phi_2) = (0^\circ, 0^\circ)$ in the stream coordinates.
Black arrows originating from the progenitors indicate the direction of the stream's motion, pointing along the leading tail, which extends toward negative $\phi_1$, while the trailing tail extends toward positive $\phi_1$.
The semi-transparent black bands with central black curves represent the stream track and transverse width, $\mu \pm \sigma$, measured from the Pal~5 member candidates reconstructed from the \citet{Xiao2025} data set; these observational constraints are shown in both the RA--Dec and $(\phi_1,\phi_2)$ panels to guide the comparison with the simulated stream morphologies.
In the bottom panel, the vertical black dashed lines mark the fiducial reliable interval of the Xiao+2025 observational track, $-14^\circ<\phi_1<17^\circ$, which is used for the quantitative track, width, and density comparisons.
}
\label{fig:spatial_morphology_stacked}
\end{figure*}


We next examine how the Galactic potential affects the projected morphology of the Palomar~5 stream, focusing on the stream extent, central track, and transverse width. For consistency with the observational profiles of \citet{Erkal2017}, all subsequent stream-coordinate comparisons in this section, including those involving the reconstructed Xiao+2025 data, are performed in the $(\phi_1,\phi_2)$ system defined by \citet{Erkal2017}. In the quantitative profile comparisons, we exclude the central progenitor region, \(|\phi_1|\leq0.42^\circ\). We first describe the morphology of the simulated debris in Galactocentric coordinates, and then compare the projected track and width with the observational measurements.

The Galactocentric \(X\)--\(Y\) and \(X\)--\(Z\) projections, shown in the first two panels of Figure~\ref{fig:spatial_morphology_stacked}, reveal that different Galactic potential models produce different stream lengths and leading--trailing asymmetries.
In the flattened-halo models, the presence and time evolution of the bar have a
strong effect on the relative extension of the two tails. In the constant-speed bar
models FHT2024 and FLHT2024, the leading tail is preferentially elongated, producing
a visibly asymmetric stream morphology in both projections. By contrast, the no-bar
control model FNHT2024 produces a shorter stream, indicating that the elongated
tails in the flattened-halo barred models are primarily associated with the rotating
bar component rather than with halo flattening alone.
The decelerating-bar model FDHT2024 shows a different type of asymmetry. Instead of
stretching the leading tail as in the constant-speed barred flattened-halo models,
the decelerating bar preferentially elongates the trailing tail. A possible reason is
that bar slowdown changes the phase at which the bar acts on the Pal~5 stream. This
phase change can lead to a different length asymmetry from the constant-pattern-speed models.

The spherical-halo barred model SHT2024 is a special case. Despite including a
rotating bar, its projected stream length remains broadly comparable to that of the
static MW2014 reference model. One possible explanation is that, in this model, the
Pal~5 orbit remains close to a low-order commensurability with the bar. As shown
above, the azimuthal and radial frequencies satisfy
\(|(\Omega_\phi-\Omega_{\rm bar})/\Omega_R|\simeq0.506\), close to a 1:2 ratio.
This near-commensurability may make the accumulated bar torque more coherent over
successive radial oscillations, thereby preventing the strong one-sided elongation
seen in the flattened-halo barred models.

In real observations, the apparent length of a stellar stream is limited by the surface density contrast of the debris relative to the background. 
Very low-density portions of the stream may exist dynamically, but they would not necessarily be detected with sufficient significance in a survey. 
To mimic this effect in a simple way, we define a mock observable stream length from the projected N-body particles. 
We divide the stream into \(\Delta\phi_1=0.1^\circ\) bins and regard a bin as detectable if it contains at least ten particles within the adopted stream-coordinate selection window. 
The observable stream extent is then defined by the continuous detectable region connected to the progenitor, with endpoints \(\phi_{1,\min}\) and \(\phi_{1,\max}\). 
The quantity \(L_{\rm stream}\) in Table~\ref{tab:length_track_summary} is the angular arc length measured along the reconstructed stream track over this continuous interval, and is therefore slightly different from the simple coordinate span \(\phi_{1,\max}-\phi_{1,\min}\).

In the \((\phi_1,\phi_2)\) panel of Figure~\ref{fig:spatial_morphology_stacked}, the MW2014 and FNHT2024 models show relatively extended and continuous debris distributions. 
Since these models do not include a rotating bar, the stream particles are not strongly redistributed into localized overdensities by bar-driven perturbations. 
As a result, their detectable extents remain large, with \(L_{\rm stream}=45.2^\circ\) for MW2014 and \(42.7^\circ\) for FNHT2024.

By contrast, the FHT2024 and FLHT2024 models have shorter mock observable lengths, 
\(L_{\rm stream}=25.8^\circ\) and \(28.0^\circ\), respectively. 
Although these models produce prominent morphological distortions, Figure~\ref{fig:density_profile} shows that a substantial fraction of the debris is concentrated into an overdense region near the progenitor, while the outer stream has a lower surface density. 
Consequently, the low-density outer portions fail the detectability criterion in our mock length measurement, leading to a shorter observed stream length.

The FDHT2024 model behaves differently. 
Unlike FHT2024 and FLHT2024, it does not produce a strong overdensity concentrated near the progenitor. 
Instead, the leading side contains a compact clump around \(\phi_1\simeq -15^\circ\), which keeps this side detectable out to \(\phi_{1,\min}\simeq -19.8^\circ\). 
At the same time, the trailing tail is strongly extended by the decelerating bar and maintains a relatively continuous density distribution, reaching \(\phi_{1,\max}\simeq 23.7^\circ\). 
This gives the largest mock observable stream length among the tested models, \(L_{\rm stream}=48.4^\circ\). 
Thus, the measured stream length is controlled not only by the total spatial extent of the debris, but also by how the Galactic potential redistributes particles along the stream.

The observational stream track shown in the RA--Dec and \((\phi_1,\phi_2)\) panels of Figure~\ref{fig:spatial_morphology_stacked} is derived from our reconstructed observational measurements, with the semi-transparent band indicating the measured transverse width around the central track. At the two ends of the reconstructed profile, the fitted track shows a noticeable edge-related bending, especially for \(\phi_1<-14^\circ\) and \(\phi_1>17^\circ\). These outer regions are more sensitive to the declining surface density of the stream and to background contamination, and we therefore do not use them as robust constraints on the stream locus. In the following quantitative comparison, the track residuals are computed only over the fiducial reliable interval \(-14^\circ<\phi_1<17^\circ\).

The resulting track-matching statistics are listed in Table~\ref{tab:length_track_summary}. Among the tested models, SHT2024 shows the largest deviation from the observational track, with \(\mathrm{RMS}_{\rm track}=0.562^\circ\). This offset is also visible in the RA--Dec and \((\phi_1,\phi_2)\) projections of Figure~\ref{fig:spatial_morphology_stacked}, where the simulated stream locus is systematically displaced relative to the observed track. By contrast, the other models have smaller track residuals, typically at the level of \(\sim0.3^\circ\). In our simulations, the flattened-halo models therefore provide a somewhat closer match to the observed stream track. This sensitivity of the projected stream locus to the halo shape is consistent with \citet{Bovy2016}, who showed that changing the halo axis ratio alters the stream track on the sky by changing the relative radial and vertical components of the Galactic force field. 

We next quantify the transverse width of the simulated streams in Table~\ref{tab:width_summary}. The raw median width is the median of the binned N-body width profile measured directly from the unscaled projected particle distribution. The raw leading and trailing widths are defined in the same way, but using only the negative- and positive-\(\phi_1\) sides, respectively. These three quantities therefore describe the absolute width scale of the simulated streams.

A common trend in the simulations is that the leading tail is broader than the trailing tail. This is likely dominated, at least in part, by projection effects in the adopted stream-coordinate system. As shown in Figure~\ref{fig:spatial_morphology_stacked}, the leading side of the simulated streams generally has a stronger downward tilt in the $(\phi_1,\phi_2)$ projection, which increases the apparent transverse width when the width is measured in fixed $\phi_1$ bins. This effect is particularly clear for SHT2024: its leading tail is strongly tilted relative to the observed stream track, while the trailing side remains comparatively flat, producing a large difference between the raw leading width ($0.300^\circ$) and the raw trailing width ($0.129^\circ$). The flattened-halo models FHT2024, FNHT2024, and FLHT2024 have larger overall median widths, $0.250^\circ$, $0.266^\circ$, and $0.234^\circ$, respectively. A plausible contributing factor is the enhanced orbital-plane precession in the oblate halo: because the total angular-momentum vector is no longer conserved, the orbital plane undergoes nodal precession at a rate that depends on each star's energy and angular momentum, so that the finite spread of these quantities across the tidal debris can translate into an additional transverse spreading of the stream, an effect that has been shown to make the Pal~5 stream morphology sensitive to the geometry of the dark-matter halo \citep{Pearson2015}. This interpretation is qualitatively consistent with the stronger orbital precession found for these flattened-halo models (Figure~\ref{fig:pal5_orbit_comparison}) and with the comparatively narrow streams of the spherical-halo models, in which the orbital plane is better preserved. We caution, however, that halo flattening alone does not guarantee a larger width, as illustrated by FDHT2024 below, so we regard this as a likely rather than a definitive explanation. By contrast, FDHT2024 has a relatively small raw median width ($0.167^\circ$). Although this model forms a compact structure on the leading side, this localized feature does not dominate the median width because most of the remaining stream is comparatively narrow.

\begin{figure}
    \centering
    \includegraphics[width=\textwidth]{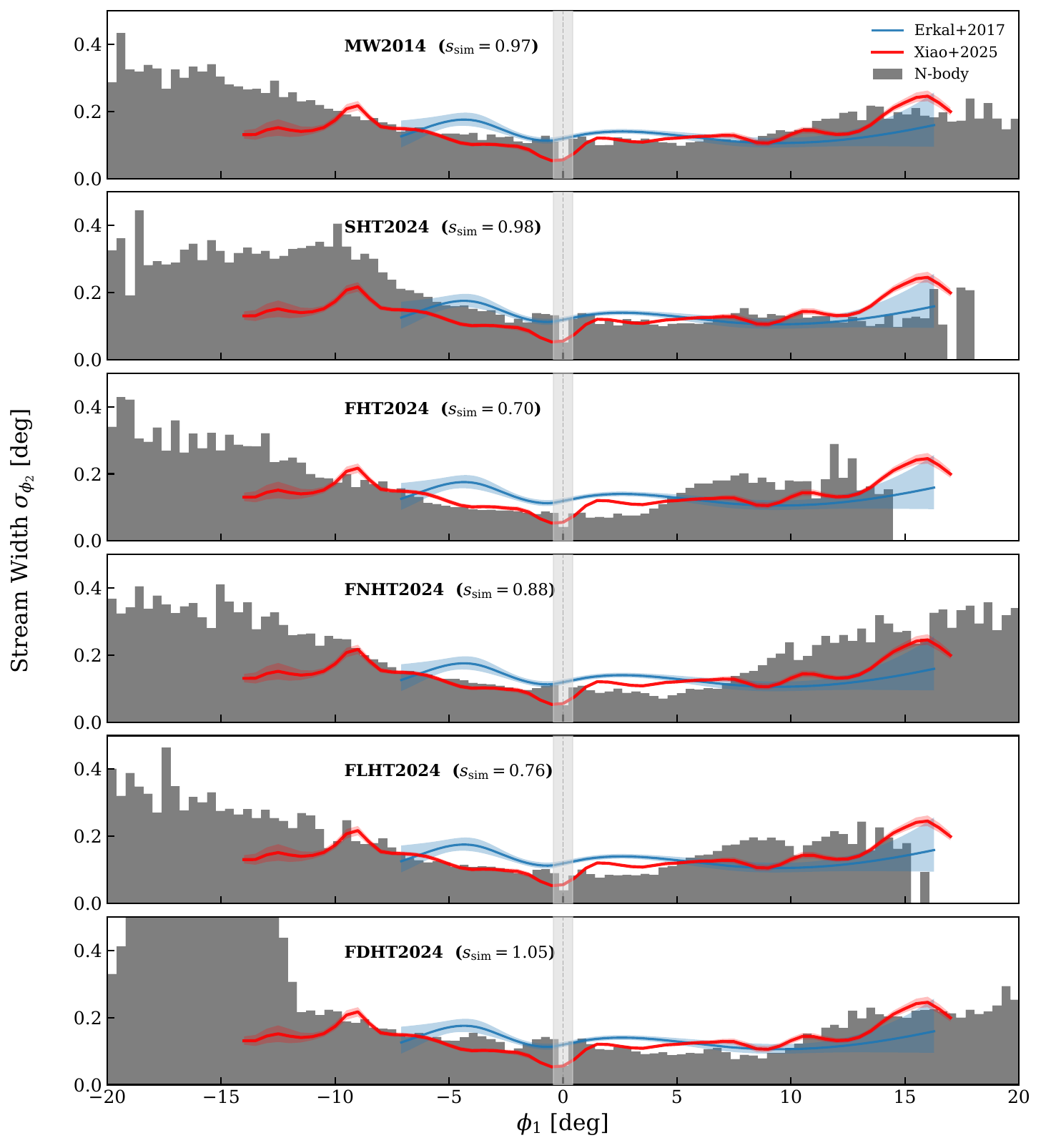}
    \caption{Comparison of the scaled stream-width profiles of the Palomar~5 stream between observations and simulations. In each panel, the grey histogram represents the scaled N-body stream width derived from this work. The red line with shaded error bands shows the stream width profile derived from the high-probability members identified in \citet{Xiao2025}. For reference, the width profile from \citet{Erkal2017} is plotted as a blue line. The simulated widths are scaled to match the observed median width outside the masked progenitor region, allowing a comparison of their shapes. The vertical dashed line at $\phi_1 = 0^\circ$ marks the position of the progenitor cluster. The grey shaded region indicates the masked area, which is excluded from the scaling calculation.}
    \label{fig:width_profile}
\end{figure}

For the profile-level comparison in Figure~\ref{fig:width_profile}, the simulated width profiles are normalized before computing the RMS residuals. Specifically, each N-body width profile is multiplied by a factor \(s_{\rm sim}\), chosen to match the median width of \citet{Erkal2017} over the common comparison range. The value of \(s_{\rm sim}\) is indicated in each panel of Figure~\ref{fig:width_profile}. The reconstructed Xiao+2025 width profile is also normalized to the \citet{Erkal2017} width scale for shape comparison, while the Erkal+2017 profile itself is left unchanged. RMS$_{\rm Xiao}$ is computed over the reliable Xiao+2025 interval \(-14^\circ<\phi_1<17^\circ\), whereas RMS$_{\rm Erkal}$ is computed over the Erkal+2017 interval \(-7^\circ<\phi_1<16^\circ\). The SHT2024 model has a relatively large RMS$_{\rm Xiao}$ because its leading tail is too broad compared with the Xiao+2025 profile. The largest RMS$_{\rm Xiao}$ occurs for FDHT2024, mainly because the compact structure near \(\phi_1\simeq -13^\circ\) produces an extreme local width that increases the profile residual. For the \citet{Erkal2017} comparison, the observed width profile varies more smoothly; the MW2014 and SHT2024 models follow this behaviour most closely and therefore have the smallest RMS$_{\rm Erkal}$ values. In contrast, FNHT2024 shows a more rapid increase in width across the comparison range, giving the largest RMS$_{\rm Erkal}$ among the tested models.


\begin{table*}[htbp]
\centering
\caption{Projected stream lengths and track residuals for the simulated Palomar~5 streams.}
\label{tab:length_track_summary}
\begin{tabular}{lcccc}
\hline
\hline
Potential Model & $\phi_{1,\min}$ & $\phi_{1,\max}$ & $L_{\rm stream}$ & RMS$_{\rm track}$ \\
 & (deg) & (deg) & (deg) & (deg) \\
\hline
MW2014   & $-22.9$ & 19.7 & 45.2 & 0.319 \\
SHT2024  & $-17.2$ & 15.5 & 34.4 & 0.562 \\
FHT2024  & $-13.2$ & 11.7 & 25.8 & 0.317 \\
FNHT2024 & $-21.4$ & 18.4 & 42.7 & 0.286 \\
FLHT2024 & $-14.3$ & 12.7 & 28.0 & 0.289 \\
FDHT2024 & $-19.8$ & 23.7 & 48.4 & 0.261 \\
\hline
\end{tabular}
\end{table*}


\begin{table*}[htbp]
\centering
\caption{Quantitative comparison of the Palomar~5 stream width profiles. The median, leading-side, and trailing-side widths are measured directly from the unscaled projected N-body particles using the ridge-based width definition. The RMS values are computed after normalizing the simulated width profiles to the median width of \citet{Erkal2017}, and therefore quantify the profile-shape mismatch rather than the absolute width scale.}
\label{tab:width_summary}
\begin{tabular}{lccccc}
\hline
\hline
Potential Model & Raw Median Width & Raw Leading Width & Raw Trailing Width & RMS$_{\rm Xiao}$ & RMS$_{\rm Erkal}$ \\
 & (deg) & (deg) & (deg) & (deg) & (deg) \\
\hline
MW2014   & 0.179 & 0.215 & 0.150 & 0.047 & 0.038 \\
SHT2024  & 0.151 & 0.300 & 0.129 & 0.089 & 0.026 \\
FHT2024  & 0.250 & 0.272 & 0.228 & 0.058 & 0.062 \\
FNHT2024 & 0.266 & 0.283 & 0.234 & 0.075 & 0.080 \\
FLHT2024 & 0.234 & 0.294 & 0.213 & 0.057 & 0.057 \\
FDHT2024 & 0.167 & 0.207 & 0.132 & 0.152 & 0.042 \\
\hline
\end{tabular}
\end{table*}


\subsubsection{Linear density}
\label{sec:stream_density}


\begin{figure}
    \centering
    \includegraphics[width=\textwidth]{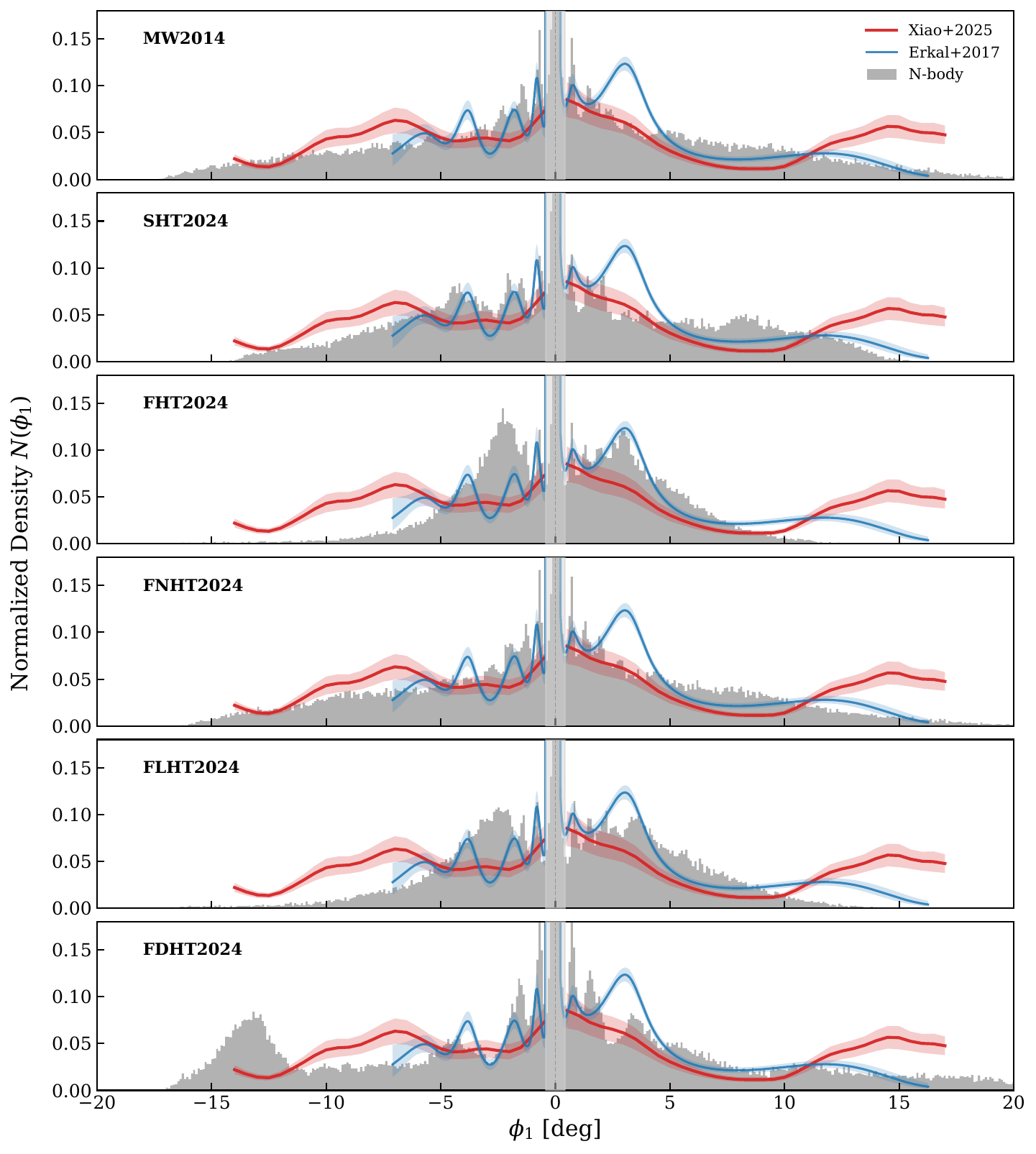}
    \caption{Comparison of the normalized linear density profiles of the Palomar~5 stream between observations and simulations. In each panel, the grey histogram represents the simulated stream profile derived from this work. The red line with shaded error bands shows the stream line-density profile derived from the high-probability members identified in \citet{Xiao2025}. For reference, the profile from \citet{Erkal2017} is plotted as a blue line. The profiles are normalized to compare their shapes. The vertical dashed line at $\phi_1 = 0^\circ$ marks the position of the progenitor cluster. The grey shaded region indicates the masked area, which is excluded from the normalization calculation.}
    \label{fig:density_profile}
\end{figure}


Figure~\ref{fig:density_profile} compares the normalized line-density profiles of the observed and simulated Palomar~5 streams. The two models without a Galactic bar, MW2014 and FNHT2024, show relatively smooth density profiles, with no strong off-centre overdensity over most of the plotted range. By contrast, the two flattened-halo models with a constant-speed bar, FHT2024 and FLHT2024, display similar density structures, with enhanced densities on both sides of the progenitor, approximately around \(\phi_1\simeq \pm3^\circ\). Although SHT2024 also includes a constant-speed bar, its density profile differs from those of FHT2024 and FLHT2024, likely because the Pal~5 orbit in this spherical halo model is affected by the near-commensurability with the bar discussed above. The FDHT2024 model shows a distinct asymmetric density profile, with a prominent overdensity near \(\phi_1\simeq -13^\circ\), corresponding to the compact structure seen in the stream morphology. These model-to-model differences indicate that the Galactic bar can strongly redistribute debris along the Pal~5 stream. In addition, all models produce narrow density enhancements close to the progenitor. Following the interpretation of \citet{Erkal2017}, these inner peaks may be related to epicyclic overdensities associated with recently stripped stars.

The comparison with the observational profiles shows that no single model reproduces all of the density features. The two observational profiles themselves also differ in several regions. Both data sets show a low-density region near $\phi_1\simeq -3^\circ$, which is opposite to the enhanced inner densities produced by the FHT2024 and FLHT2024 models. On the leading side, the reconstructed Xiao+2025 profile shows an overdensity near $\phi_1\simeq 15^\circ$, while the \citet{Erkal2017} profile shows only a mild enhancement around $\phi_1\simeq 12^\circ$ and then decreases toward the outer edge of its coverage. In addition, the \citet{Erkal2017} profile shows a clear overdensity near $\phi_1\simeq 3^\circ$, which is qualitatively similar to the features in FHT2024 and FLHT2024, whereas the reconstructed Xiao+2025 profile does not show an equally prominent overdensity at this location. In the interval $-10^\circ \lesssim \phi_1 \lesssim -5^\circ$, the Xiao+2025 profile shows relatively high density. Some models, such as MW2014, SHT2024, and FNHT2024, also maintain comparatively high density in this region, but none produces a distinct overdensity that matches the observed structure.

These differences between the two observational profiles may reflect the different survey data sets and analysis pipelines: the Xiao+2025 reconstruction is based on the DESI Legacy Imaging Surveys DR10 photometry, whereas \citet{Erkal2017} used SDSS data. The deeper imaging, different footprint, member selection, and background treatment can all affect the recovery of low-contrast density features in the outer stream. Overall, the density comparison provides a useful diagnostic of how different Galactic potentials redistribute stream debris, but the current set of models does not provide a complete match to the observed Pal~5 line-density profile. In particular, none of the simulated models produces a clear overdensity at the same location, $\phi_1\simeq 15^\circ$, as the reconstructed Xiao+2025 feature.


\section{Discussion and Conclusions} \label{sec:discussion_conclusion}


In this paper, we investigated the recent dynamical evolution of the Pal~5 globular
cluster and its tidal stream using direct \(N\)-body simulations with \texttt{PeTar}.
By comparing six Galactic potential models, we examined how the LMC, halo shape,
Galactic bar, spiral arms, and bar deceleration affect the bound cluster and its
tidal debris.

The comparison between these simulations and the observational measurements shows
that different Galactic components affect the Pal~5 stream in distinct ways. The
Galactic bar has a clear impact on the stream length and on the redistribution of
debris along the tails, consistent with previous studies showing that
non-axisymmetric perturbations can modify tidal-stream morphologies, density
structures, and leading--trailing asymmetries
\citep[e.g.,][]{Hattori2016, PriceWhelan2016, Erkal2017}. The halo shape mainly
affects the projected stream track. This is consistent with the conclusion of
\citet{Bovy2016} that the Pal~5 stream is highly sensitive to changes in the halo
axis ratio and in the flattening of the Galactic force field. In our simulations,
the halo shape also changes the vertical oscillation frequency of the Pal~5 orbit,
which in turn modifies its resonant interaction with the Galactic bar. The LMC produces only a modest
direct change in the present-day projected morphology of the stream in our adopted
models, but it can change the pericentric distance and therefore affect the mass
evolution of the progenitor.

Although these models reproduce some individual properties of the stream, no single
model in our suite simultaneously matches all observed features in the stream
length, track, width, and line-density profile. This highlights the complexity of the Milky Way's evolution: in addition to the bar, spiral structure, halo shape, and the LMC, effects such as the response of the halo to infalling satellites, an
evolving or tilted disk, small-scale perturbers, and observational selection can also
leave signatures in stellar streams
\citep[e.g.,][]{Pearson2015, Erkal2019, GaravitoCamargo2019, Vasiliev2021,
Shipp2021, Nibauer2024}.

Several limitations should be kept in mind when interpreting these results. First, our model grid is discrete and was designed to isolate the effects of selected Galactic components rather than to perform a full parameter search. We do not explore the full range of halo flattening, bar pattern speed, bar deceleration rate, spiral-arm parameters, or LMC properties. We also include spiral arms in several models, but do not separately isolate their dynamical effects or vary their phase and pattern speed. A broader search over this parameter space is needed before drawing stronger constraints on the Milky Way potential. Because direct \(N\)-body simulations are computationally expensive, a restricted \(N\)-body approach may provide a useful intermediate step between particle-spray methods and full \(N\)-body calculations. Such models can follow the internal evolution of the progenitor and release stream particles at a self-consistently determined rate, while remaining computationally cheaper than a large grid of full \(N\)-body simulations \citep{Vasiliev2021,Limberg2025}. Our models can also serve as benchmarks for calibrating restricted and particle-spray methods.

Second, we fix both the present-day phase-space coordinates and the internal model of the Pal~5 progenitor. Uncertainties in the current position and velocity of the cluster could shift the projected stream track and affect the inferred stream morphology. In addition, varying the initial cluster structure, such as the concentration, density profile, mass function, or the choice of King-model parameters, may change the stripping history and enhance or suppress some small-scale stream features. These progenitor-model uncertainties are not explored in the present work.

Third, we integrate the cluster over only the last \(\sim3~{\rm Gyr}\). 
This choice is motivated not only by computational cost, but also by the limited
constraints on the detailed structure and time evolution of the Milky Way potential
over cosmological timescales. 
Restricting the simulations to the recent evolution reduces the risk that small
uncertainties in the assumed Galactic potential accumulate into unrealistically
large orbital errors over long integrations. 

Finally, our simulations use smooth analytic Galactic potentials and therefore omit small-scale perturbers, such as dark matter subhalos, giant molecular clouds, and other clusters, which can contribute to local density variations and dynamical heating of the stream.

Looking forward, in order to distinguish between the various dynamical mechanisms, whether the stream features are caused by the rotating bar, impacts by giant molecular clouds, or dark matter subhalos, future work should explore a broader parameter space and, where the past evolution of the Milky Way potential can be constrained, test longer integrations with more
realistic time-dependent models. Furthermore, with the advent of future large-scale survey projects such as LSST and CSST, we will obtain deeper and wider observational data on stellar streams. By combining these high-quality data with next-generation simulations that incorporate these small-scale physical processes, it should be possible to identify the specific dynamical drivers responsible for the detailed morphology of the Pal~5 stream.


\begin{acknowledgments}
We thank the anonymous referee for constructive comments that improved this manuscript. We thank Professor Xiao for providing the pre-processed DESI Legacy Imaging Surveys DR10 photometric catalogue used in this work. LW acknowledges support from the National Natural Science Foundation of China through grants 12573041 and 12233013, the High-level Youth Talent Project (Provincial Financial Allocation) through grant 2023HYSPT0706, and the Fundamental Research Funds for the Central Universities, Sun Yat-sen University (2025QNPY04). EV acknowledges support from an STFC Ernest Rutherford fellowship (ST/X004066/1).
The authors acknowledge the Beijing Super Cloud Center for providing HPC resources that have contributed to the research results reported within this paper. URL: http://www.blsc.cn/.
\end{acknowledgments}

\begin{contribution}
ZH was responsible for the method development, performing the simulations, data analysis, and writing the manuscript.
LW came up with the initial research concept, contributed to discussions, edited the manuscript, and supervised the overall project.
ZZ contributed to the Galactic-potential method development and discussions.
YH provided observational data, contributed to discussions and edited the manuscript.
EV helped on the implementation of \texttt{AGAMA} in \texttt{PeTar} code, contributed to discussions and edited the manuscript.

\end{contribution}


\appendix


\section{Basis Potential Parameters}
\label{app:potential_parameters}


This appendix summarizes the main parameters of the two basis Galactic potentials
used in this work. The first is the time-independent \texttt{MWPotential2014}
reference model of \citet{2015ApJS..216....9B}, whose main component parameters are
listed in Table~\ref{tab:mw2014_parameters}. The second is the
\citet{hunterTestingKinematicDistances2024}-based Milky Way model implemented with
\texttt{AGAMA}, whose main component parameters are listed in
Table~\ref{tab:ht2024_parameters}. The individual simulation models discussed in the
main text (MW2014, SHT2024, FHT2024, FNHT2024, FLHT2024, and FDHT2024) are
constructed by adopting, modifying, or combining these basis components as described
in Section~\ref{sec:potential}. Therefore, Tables~\ref{tab:mw2014_parameters} and \ref{tab:ht2024_parameters} list the component parameters of the underlying basis potentials rather than repeating the full definition of each model variant.

\begin{table*}[htbp]
\centering
\caption{Main parameters of the \texttt{MWPotential2014} reference potential.}
\label{tab:mw2014_parameters}
\begin{tabular}{llll}
\hline
\hline
Component & Functional Form & Parameter & Value \\
\hline
Dark matter halo & Spherical NFW halo
& Virial mass, $M_{\rm vir}$ & $8.0\times10^{11}\,M_\odot$ \\
& & Concentration, $c$ & $15.3$ \\
& & Scale radius, $r_s$ & $16\,{\rm kpc}$ \\
\hline
Disk & Miyamoto--Nagai disk
& Mass, $M$ & $6.8\times10^{10}\,M_\odot$ \\
& & Scale length, $a$ & $3\,{\rm kpc}$ \\
& & Scale height, $b$ & $280\,{\rm pc}$ \\
\hline
Bulge & Power-law density with exponential cutoff
& Mass, $M$ & $5.0\times10^9\,M_\odot$ \\
& & Power-law exponent, $\alpha$ & $-1.8$ \\
& & Cutoff radius, $r_c$ & $1.9\,{\rm kpc}$ \\
\hline
\end{tabular}
\end{table*}

\begin{table*}[!t]
\centering
\caption{Main parameters of the \citet{hunterTestingKinematicDistances2024}-based basis potential.}
\label{tab:ht2024_parameters}
\begin{tabular}{llll}
\hline
\hline
Component & Description & Parameter & Value \\
\hline
Central black hole & Sgr~A$^\ast$ Plummer potential
& Mass, $M_{\rm Sgr\,A^\ast}$ & $4.154\times10^6\,M_\odot$ \\
& & Softening scale, $b$ & $0.1\,{\rm pc}$ \\
\hline
Nuclear star cluster & Flattened Dehnen profile
& Mass, $M_{\rm NSC}$ & $6.1\times10^7\,M_\odot$ \\
& & Scale radius, $a_0$ & $5.9\,{\rm pc}$ \\
& & Inner slope, $\gamma$ & $0.71$ \\
& & Flattening, $q$ & $0.73$ \\
\hline
Nuclear stellar disk & Double exponential-like NSD
& $R_1$, $R_2$ & $5.06\,{\rm pc}$, $24.6\,{\rm pc}$ \\
& & $n_1$, $n_2$ & $0.72$, $0.79$ \\
& & Flattening, $q$ & $0.37$ \\
& & Density ratio, $\rho_1/\rho_2$ & $1.311$ \\
& & Normalization, $\rho_2$ & $1.53\times10^{12}\,M_\odot\,{\rm kpc}^{-3}$ \\
\hline
Galactic bar & Box/peanut bulge + long bar
& Total mass, $M_{\rm bar}$ & $1.83\times10^{10}\,M_\odot$ \\
& & Pattern speed, $\Omega_{\rm bar}$ &
$-37.5\,{\rm km\,s^{-1}\,kpc^{-1}}$ \\
\hline
Stellar disks & Two exponential stellar disks
& $\Sigma_1$, $R_{d,1}$, $z_1$ &
$1.3719\times10^3\,M_\odot\,{\rm pc}^{-2}$, $2.0\,{\rm kpc}$, $0.3\,{\rm kpc}$ \\
& & $\Sigma_2$, $R_{d,2}$, $z_2$ &
$9.2391\times10^2\,M_\odot\,{\rm pc}^{-2}$, $2.8\,{\rm kpc}$, $0.9\,{\rm kpc}$ \\
& & Inner cutoff, $R_{\rm cut}$ & $2.4\,{\rm kpc}$ \\
\hline
Gas disks & H{\sc i} and H$_2$ disks
& H{\sc i}: $\Sigma$, $R_d$, $z$, $R_m$ &
$53.1\,M_\odot\,{\rm pc}^{-2}$, $7.0\,{\rm kpc}$, $85\,{\rm pc}$, $4.0\,{\rm kpc}$ \\
& & H$_2$: $\Sigma$, $R_d$, $z$, $R_m$ &
$2.18\times10^3\,M_\odot\,{\rm pc}^{-2}$, $1.5\,{\rm kpc}$, $45\,{\rm pc}$, $12.0\,{\rm kpc}$ \\
\hline
Spiral arms & Stellar-disk perturbation
& Relative amplitude, $\alpha$ & $0.36$ \\
& & Pitch angle, $i$ & $12.5^\circ$ \\
& & Reference radius, $R_a$ & $9.64\,{\rm kpc}$ \\
& & Width parameter, $\sigma_{\rm sp}$ & $5\,{\rm kpc}$ \\
& & Phase angles, $\gamma_1$, $\gamma_2$ & $139.5^\circ$, $69.75^\circ$ \\
& & Pattern speed, $\Omega_{\rm spiral}$ &
$-22.5\,{\rm km\,s^{-1}\,kpc^{-1}}$ \\
\hline
Dark matter halo & Spherical Einasto halo
& Total mass, $M_{\rm halo}$ & $1.1\times10^{12}\,M_\odot$ \\
& & Einasto index, $n$ & $4.5$ \\
& & Half-mass radius, $r_s$ & $96\,{\rm kpc}$ \\
& & Scale parameter, $a$ & $0.88\,{\rm pc}$ \\
\hline
\end{tabular}
\end{table*}


\section{Verification of the 2:1 Resonant Capture}
\label{sec:appendix_resonance}

\begin{figure}[htbp]
    \centering
    \includegraphics[width=0.8\textwidth]{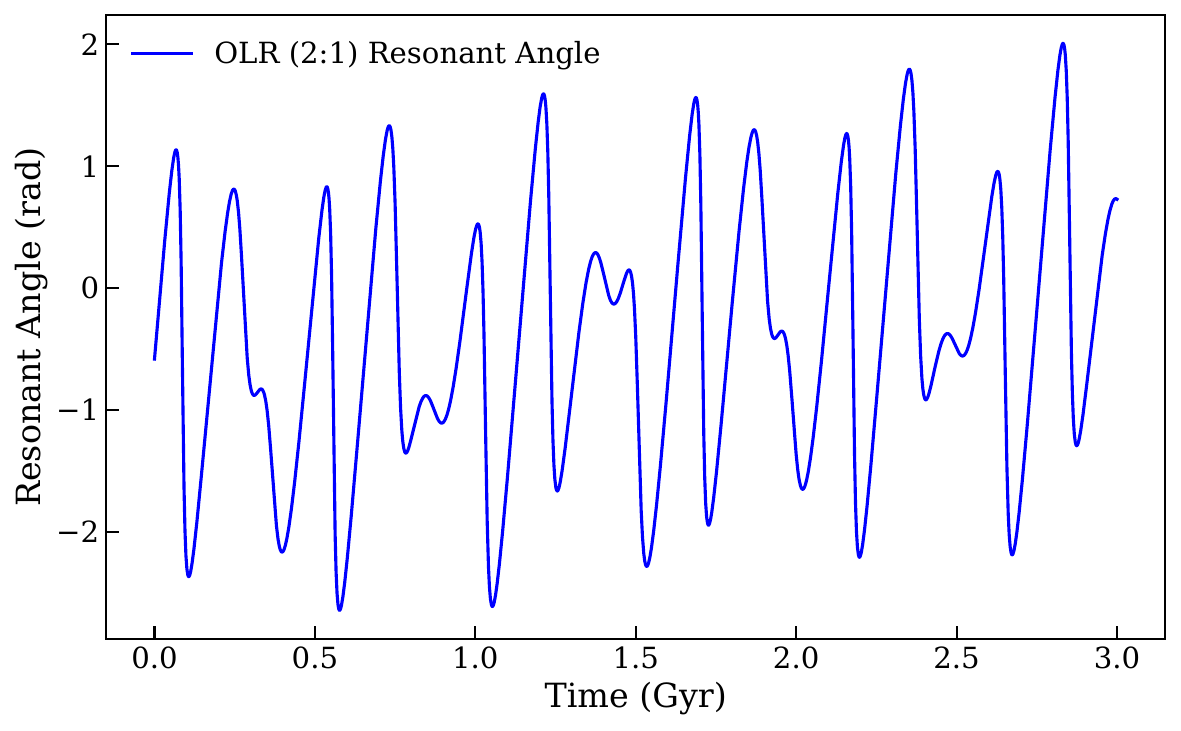}
    \caption{
        Time evolution of the 2:1 Outer Lindblad Resonance (OLR) angle for the Pal~5 progenitor in the SHT2024 potential. The resonant angle remains bounded over the \(3~\mathrm{Gyr}\) integration, indicating sustained libration rather than circulation.
    }
    \label{fig:libration_plot}
\end{figure}

To test whether the macroscopic orbital behaviour in the SHT2024 model is associated with a bar-related resonance, we evaluate the 2:1 Outer Lindblad Resonance (OLR) angle, \(\theta_{\rm res}=2(\phi-\Omega_{\rm bar}t)-\theta_R\). As shown in Figure~\ref{fig:libration_plot}, this angle does not circulate through the full \(2\pi\) range, but instead remains bounded throughout the integration. This behaviour is consistent with the Pal~5 progenitor orbit being affected by, or temporarily trapped near, a 2:1 resonance with the Galactic bar in the spherical-halo SHT2024 potential. 

For the subsequent models incorporating a flattened dark matter halo (e.g., FHT2024), the altered potential changes the intrinsic radial frequency $\Omega_R$. A similar frequency analysis reveals that the frequency ratio significantly deviates from the 2:1 commensurability, and the corresponding angle transitions into circulation. This validates our morphological observation that the flattened halo effectively breaks the resonant phase-locking, leading to the pronounced orbital precession seen in the bar-corotating frame (bottom row of Figure~\ref{fig:pal5_orbit_comparison}).


\clearpage

\bibliography{reference}{}
\bibliographystyle{aasjournalv7}

\end{document}